\documentclass[11pt]{article}\usepackage[]{graphicx}\usepackage[]{xcolor}
\makeatletter
\def\maxwidth{ %
  \ifdim\Gin@nat@width>\linewidth
    \linewidth
  \else
    \Gin@nat@width
  \fi
}
\makeatother

\definecolor{fgcolor}{rgb}{0.345, 0.345, 0.345}
\newcommand{\hlnum}[1]{\textcolor[rgb]{0.686,0.059,0.569}{#1}}%
\newcommand{\hlsng}[1]{\textcolor[rgb]{0.192,0.494,0.8}{#1}}%
\newcommand{\hlcom}[1]{\textcolor[rgb]{0.678,0.584,0.686}{\textit{#1}}}%
\newcommand{\hlopt}[1]{\textcolor[rgb]{0,0,0}{#1}}%
\newcommand{\hldef}[1]{\textcolor[rgb]{0.345,0.345,0.345}{#1}}%
\newcommand{\hlkwa}[1]{\textcolor[rgb]{0.161,0.373,0.58}{\textbf{#1}}}%
\newcommand{\hlkwb}[1]{\textcolor[rgb]{0.69,0.353,0.396}{#1}}%
\newcommand{\hlkwc}[1]{\textcolor[rgb]{0.333,0.667,0.333}{#1}}%
\newcommand{\hlkwd}[1]{\textcolor[rgb]{0.737,0.353,0.396}{\textbf{#1}}}%

\usepackage{framed}
\makeatletter
\newenvironment{kframe}{%
 \def\at@end@of@kframe{}%
 \ifinner\ifhmode%
  \def\at@end@of@kframe{\end{minipage}}%
  \begin{minipage}{\columnwidth}%
 \fi\fi%
 \def\FrameCommand##1{\hskip\@totalleftmargin \hskip-\fboxsep
 \colorbox{shadecolor}{##1}\hskip-\fboxsep
     \hskip-\linewidth \hskip-\@totalleftmargin \hskip\columnwidth}%
 \MakeFramed {\advance\hsize-\width
   \@totalleftmargin\z@ \linewidth\hsize
   \@setminipage}}%
 {\par\unskip\endMakeFramed%
 \at@end@of@kframe}
\makeatother

\definecolor{shadecolor}{rgb}{.97, .97, .97}
\definecolor{messagecolor}{rgb}{0, 0, 0}
\definecolor{warningcolor}{rgb}{1, 0, 1}
\definecolor{errorcolor}{rgb}{1, 0, 0}
\newenvironment{knitrout}{}{} 

\usepackage{alltt}
\usepackage[margin=1in]{geometry}
\usepackage{amsmath, amsfonts, amssymb}
\usepackage{bm}
\usepackage{url}
\usepackage{natbib}

\usepackage{parskip}
\usepackage[bf,font={sl}]{caption}
\usepackage{bm}

\let\code=\texttt
\let\proglang=\textsf
\newcommand{\pkg}[1]{{\fontseries{b}\selectfont #1}}

\author{Théo Michelot\footnote{theo.michelot@dal.ca}\\Dalhousie University}
\title{\pkg{hmmTMB}: hidden Markov models with flexible\\ covariate effects in \proglang{R}}
\IfFileExists{upquote.sty}{\usepackage{upquote}}{}
\begin{document}
\maketitle

\begin{abstract}
  Hidden Markov models (HMMs) are widely applied in studies where a discrete-valued process of interest is observed indirectly. They have for example been used to model behaviour from human and animal tracking data, disease status from medical data, and financial market volatility from stock prices. The model has two main sets of parameters: transition probabilities, which drive the latent state process, and observation parameters, which characterise the state-dependent distributions of observed variables. One particularly useful extension of HMMs is the inclusion of covariates on those parameters, to investigate the drivers of state transitions or to implement Markov-switching regression models. We present the new \proglang{R} package \pkg{hmmTMB} for HMM analyses, with flexible covariate models in both the hidden state and observation parameters. In particular, non-linear effects are implemented using penalised splines, including multiple univariate and multivariate splines, with automatic smoothness selection. The package allows for various random effect formulations (including random intercepts and slopes), to capture between-group heterogeneity. \pkg{hmmTMB} can be applied to multivariate observations, and it accommodates various types of response data, including continuous (bounded or not), discrete, and binary variables. Parameter constraints can be used to implement non-standard dependence structures, such as semi-Markov, higher-order Markov, and autoregressive models. Here, we summarise the relevant statistical methodology, we describe the structure of the package, and we present an example analysis of animal tracking data to showcase the workflow of the package.
\end{abstract}

\section{Introduction}

Hidden Markov models (HMMs) have been applied in many areas, including medicine \citep{altman2005}, ecology \citep{mcclintock2020}, and finance \citep{bulla2006}. They are useful when an observed phenomenon is driven by several unobserved regimes, or states, and when it is of interest to model the state-switching dynamics. We first describe the basic mathematical formulation of discrete-time HMMs, and we refer the reader to \cite{zucchini2017} for more detail. 

An HMM consists of two stochastic processes: a state process $(S_t)$, which can take on a finite number of values $\{ 1, 2, \dots, K \}$, and an observation process $(Z_t)$. At each time step, $Z_t$ is assumed to depend on the current state $S_t$ through a state-dependent distribution,
\begin{equation*}
  p(Z_t \mid Z_1, \dots, Z_{t-1}, S_1, \dots, S_{t-1}, S_t) = p(Z_t \mid S_t).
\end{equation*}

The observation distribution (also called ``emission distribution'') is often taken from some parametric family, where the parameters are state-dependent, i.e., $Z_t \mid \{ S_t = j \} \sim \mathcal{D}(\bm\omega_j)$ for $j \in \{1, \dots, K\}$. For example, if $\mathcal{D}$ is the normal distribution, $\bm\omega_j$ might be a vector of the mean and standard deviation of $Z_t$ in state $j$. In the case where there are several observed variables $\bm{Z}_t = (Z_{1t}, Z_{2t}, \dots, Z_{dt})$, it is most common to assume that they are conditionally independent given the state, i.e., $p(\bm{Z}_t \mid S_t) = p(Z_{1t} \mid S_t) \times p(Z_{2t} \mid S_t) \times \cdots \times p(Z_{dt} \mid S_t)$. This assumption of contemporaneous conditional independence is often reasonable because dependence between the variables is induced by the state process.

The state process is specified as a first-order Markov chain, i.e.,
\begin{equation}
  \label{eqn:markov}
  p(S_t \mid S_1, \dots, S_{t-1}) = p(S_t \mid S_{t-1}).
\end{equation}

It is parameterised in terms of an $K \times K$ transition probability matrix
\begin{equation*}
  \bm\Gamma =
  \begin{pmatrix}
    \gamma_{11} & \gamma_{12} & \cdots & \gamma_{1K} \\
    \gamma_{21} & \gamma_{22} & \cdots & \gamma_{2K} \\
    \vdots & \vdots & \ddots & \vdots \\
    \gamma_{K1} & \gamma_{K2} & \cdots & \gamma_{KK}
  \end{pmatrix}
\end{equation*}
where $\gamma_{ij} = \Pr(S_t = j \mid S_{t-1} = i)$, subject to the constraints that elements from each row sum to 1, i.e., $\sum_{k=1}^K \gamma_{ik} = 1$ for any $i \in \{ 1, \dots, K \}$. The specification of the Markov chain also requires a vector of $K$ initial probabilities, $\bm\delta^{(1)} = (\Pr(S_1 = 1), \Pr(S_1 = 2), \dots, \Pr(S_1 = K))$, which must sum to 1.

Given observations $\bm{z}_1, \bm{z}_2, \dots, \bm{z}_n$ from the process $(Z_t)$, the problem of inference is typically to estimate all model parameters: the observation parameters $\{ \bm\omega_1, \dots, \bm\omega_K \}$, the transition probability matrix $\bm\Gamma$, and the initial distribution $\bm\delta^{(1)}$. This is a difficult problem because the state process is unobserved, but a computationally efficient procedure exists to evaluate the likelihood of the data by integrating over all possible state sequences \citep[the ``forward algorithm'';][]{zucchini2017}. The likelihood can usually be optimised numerically to obtain parameter estimates. This can be extended directly to the case of several independent time series: the full likelihood is the product of the likelihoods of the individual time series obtained from the forward algorithm. 

In this paper, we introduce the new package \pkg{hmmTMB} for the \proglang{R} environment \citep{R2022}, which provides tools to implement HMMs where parameters of interest can be specified as flexible functions of covariates, such as non-linear effects modelled using splines, and random effects. The package implements a wide variety of model formulations, including mixed HMMs \citep{altman2007}, non-homogeneous HMMs \citep{hughes1999}, and Markov-switching regression and generalised additive models \citep{kim2008, langrock2017}. We anticipate that it will be useful to many researchers in fields where HMMs are applied.

\section{Hidden Markov models with non-parametric and random effects}

The main focus of \pkg{hmmTMB} is to provide inferential tools for hidden Markov models that include flexible covariate dependences on the transition probabilities or on the state-dependent observation distribution parameters. In this section, we give a brief overview of the modelling approach and of its implementation in \pkg{hmmTMB}.

\subsection{Model formulation}
\label{sec:model}

\subsubsection{Linear predictor}

Consider a generic parameter $\theta$, which can be either a transition probability or an observation parameter, and the corresponding linear predictor $\eta = h(\theta)$, where $h$ is a link function defined so that $\eta$ is real-valued. (We discuss the choice of link function below.) To capture effects of covariates on $\theta$, we write the linear predictor using a mixed linear model approach,
\begin{equation}
  \label{eqn:linpred}
  \eta = \bm{X} \bm\alpha + \bm{R} \bm\beta,\qquad \bm\beta \sim N(\bm{0},\ \bm{Q}),
\end{equation}
where $\bm{X}$ and $\bm{R}$ are the design matrices for fixed and random effects, respectively, $\bm\alpha$ is the vector of fixed effects, and $\bm\beta$ is the vector of random effects with covariance matrix $\bm{Q}$.

Here, we use the term ``random effect'' loosely, to refer to any term that enters the model as specified above, and it encompasses not only models for inter-group variability (e.g., random intercept), but also non-parametric covariate effects based on penalised splines. This broad definition of random effects is for example discussed by \cite{hodges2013}. In the case of i.i.d.\ normal random effects, the corresponding columns of $\bm{R}$ consist of 0s and 1s, indicating which group each row belongs to, and the $\bm\beta$s are levels of the random effect (e.g., group-specific intercept), which are typically not of direct interest. In contrast, when using splines, each column of $\bm{R}$ corresponds to a basis function, where each row is its evaluation for given covariate values. The corresponding coefficients $\bm\beta$ are weights for the basis functions, which determine the shape of the resulting spline, and they are constrained by the random effect distribution to impose smoothness. For penalised splines, the value of $\bm\beta$ is of interest, as it is used to infer the shape of the non-linear relationship. The distribution of $\bm\beta$ can also be viewed as a (possibly improper) Bayesian prior, which captures the assumption that the resulting functions should be smooth \citep{wood2017, miller2025}. 

The structure of the matrix $\bm{Q}$ reflects the dependence between entries of $\bm\beta$. It is typically a block-diagonal matrix, where each block corresponds to a different random effect (or, equivalently, penalised spline) included in the model. The block for an i.i.d.\ normal random effect is diagonal, with diagonal elements equal to some variance parameter $\sigma^2$. The block for a penalised spline can be written as $\lambda^{-1} \bm{S}^-$, where $\bm{S}^-$ is the pseudo-inverse of the penalty matrix for the chosen basis \citep{wood2017}, and $\lambda > 0$ is a smoothness parameter. Note that the i.i.d.\ normal case can then be viewed as a special case of penalised spline, where $\bm{S}$ is the identity matrix, and where $\lambda = 1/\sigma^2$. In the following, we do not distinguish between variance and smoothness parameters.

An HMM includes many parameters (transition probabilities and state-dependent observation parameters), and each has an associated linear predictor as described above. There is therefore potentially one $\bm{X}$, $\bm\alpha$, $\bm{R}$, $\bm\beta$, and $\bm{Q}$ for each HMM parameter. In practice, these can be combined and stored conveniently as block-diagonal matrices.

\subsubsection{Transition probabilities}

When covariates affect the dynamics of the state process, we denote as $\bm\Gamma^{(t)}$ the transition probability matrix at time $t$, and the transition probabilities are related to the linear predictor using the multinomial logit link,
\begin{equation*}
  \gamma_{ij}^{(t)} = \frac{\exp(\eta_{ij}^{(t)})}{\sum_{k=1}^K \exp(\eta_{ik}^{(t)})},
\end{equation*}
where $\eta_{ij}^{(t)}$ is defined by Equation \ref{eqn:linpred} for $i \neq j$, with the convention that $\eta_{ii}^{(t)} = 0$. Note that there are only $K \times (K - 1)$ linear predictors for a $K$-state model, due to row constraints of the transition probability matrix. Here, we assume that the non-diagonal elements are estimated, but a different reference element could be chosen for each row. This gives rise to a non-homogeneous HMM, a useful framework to investigate the effects of covariates on the dynamics of the state process \citep{hughes1999, patterson2009}.

\subsubsection{Observation parameters}

Covariate effects can also be included in the state-dependent parameters of the observation distributions. The state-dependent distribution is then time-varying, and we write $Z_t \mid \{ S_t = j \} \sim \mathcal{D}(\omega_{j1}^{(t)}, \dots, \omega_{jL}^{(t)})$, where each $h(\bm\omega_{jl}^{(t)})$ is modelled as in Equation \ref{eqn:linpred}. In this case, the link function $h$ depends on the domain of defintion of each parameter $\omega_{jl}$; for example, an identity link might be used for a real-valued parameter, or a log link for a positive-valued parameter. This formulation includes as special case a rich family of Markov-switching regression models, including Markov-switching generalised linear and additive models \citep{kim2008, langrock2017}.

\subsection[Implementation using TMB and mgcv]{Implementation using \pkg{TMB} and \pkg{mgcv}}
\label{sec:implementation}

\subsubsection[Marginal likelihood with TMB]{Marginal likelihood with \pkg{TMB}}

Given (possibly multivariate) observations $\bm{z}_1, \bm{z}_2, \dots, \bm{z}_n$ from the process $(\bm{Z}_t)$, the primary aim of the analysis is to estimate the fixed effect parameters $\bm\alpha$ (including intercepts and linear effects), and the smoothness parameters $\bm\lambda$, which parameterise the covariance matrix $\bm{Q}$ of the random effect distribution. (A secondary aim might be to predict the random effect parameters $\bm\beta$ themselves, which we discuss later.) We propose using maximum likelihood estimation, based on the marginal likelihood of the fixed effect and smoothness parameters. 

The distribution of $\bm\beta$ (Equation \ref{eqn:linpred}) can be captured by penalising the joint log-likelihood of fixed and random effects, as,
\begin{equation*}
  l_p(\bm\alpha, \bm\beta, \bm\lambda \mid \bm{z}_1, \dots, \bm{z}_n) = \log \{ L(\bm\alpha, \bm\beta \mid \bm{z}_1, \dots, \bm{z}_n )\} - \sum_i \lambda_i \bm\beta_i^\intercal \bm{S}_i \bm\beta_i
\end{equation*}
where $L(\bm\alpha, \bm\beta \mid \bm{z}_1, \dots, \bm{z}_n)$ is the unpenalised HMM likelihood of parameters $\bm\alpha$ and $\bm\beta$ (e.g., computed using the forward algorithm), and the penalty is obtained as the log of the density function of a multivariate normal distribution (excluding additive constants). The sum is over all random effects included in the model.

For model fitting, we consider the marginal likelihood of $\bm\alpha$ and $\bm\lambda$, obtained by integrating out the random effects $\bm\beta$,
\begin{align*}
  L(\bm\alpha, \bm\lambda \mid \bm{z}_1, \dots, \bm{z}_n) & = \int \exp \{ l_p(\bm\alpha, \bm\beta, \bm\lambda \mid \bm{z}_1, \dots, \bm{z}_n )\}\ d\bm\beta \\
                                                        & = \int p(\bm{z}_1, \dots, \bm{z}_n \mid \bm\alpha, \bm\beta)\ p(\bm\beta \mid \bm\lambda)\ d\bm\beta
\end{align*}
where $p(\bm{z}_1, \dots, \bm{z}_n \mid \bm\alpha, \bm\beta) = L(\bm\alpha, \bm\beta \mid \bm{z}_1, \dots, \bm{z}_n )$ is the HMM likelihood, and $p(\bm\beta \mid \bm\lambda)$ is a multivariate normal density function with mean zero and block-diagonal precision matrix with blocks $\lambda_i \bm{S}_i$. We use the Template Model Builder (\pkg{TMB}) \proglang{R} package to compute the marginal likelihood by integrating the random effects $\bm\beta$ based on the Laplace approximation \citep{kristensen2016}. The marginal likelihood can then be optimised numerically to obtain estimates of $\bm\alpha$ and $\bm\lambda$. In cases where the levels of the random effects $\bm\beta$ are of direct interest, predicted values can be obtained, akin to best linear unbiased predictions. This is for example key for penalised splines, where these predictions are needed to derive the estimated non-linear function.

As an alternative to the marginal likelihood approach, Markov chain Monte Carlo could be used to sample over random effects in a Bayesian framework, and posterior inference could be carried out on all model parameters, as discussed in Section \ref{sec:bayes}.

Although they do not focus on mixed effect models like the ones presented in this paper, \cite{bacri2022} and \cite{bacri2023} describe in detail the implementation of basic HMMs using \pkg{TMB}. These are good additional resources for users who would like to better understand the capabilities of \pkg{TMB}, or who might want to implement HMMs from scratch.

\subsubsection[Spline specification with mgcv]{Spline specification with \pkg{mgcv}}

The approach described so far for penalised splines assumes that the design matrix of basis functions $\bm{R}$ and the penalty matrix $\bm{S}$ are known. In practice, both matrices depend on the choice of the type of spline (e.g., cubic spline, or thin plate regression spline), and on the number of basis functions. We use the \proglang{R} package \pkg{mgcv} to specify $\bm{R}$ and $\bm{S}$ \citep{wood2017}, which provides great flexibility to define a wide range of basis-penalty models.  

\subsubsection{Confidence intervals}

\pkg{TMB} can output point estimates for all model parameters (including predicted values of the random effects), that we denote as $(\hat{\bm\alpha}, \hat{\bm\beta}, \hat{\bm\lambda})$. It also computes the estimated joint precision matrix of the parameters (in \code{TMB::sdreport()}), and we can take its inverse to get the estimated covariance matrix $\hat{\bm\Sigma}$.  Following asymptotic maximum likelihood theory, the estimators approximately follow $(\hat{\bm\alpha}, \hat{\bm\beta}, \hat{\bm\lambda}) \sim N \left[ (\bm\alpha, \bm\beta, \bm\lambda),\ \hat{\bm\Sigma} \right]$. Based on this result, we apply the following simulation-based procedure to create confidence intervals:
\begin{enumerate}
\item Generate a large number $J$ of samples $(\bm\alpha^{(j)}, \bm\beta^{(j)}, \bm\lambda^{(j)}) \sim N \left[ (\hat{\bm\alpha}, \hat{\bm\beta}, \hat{\bm\lambda}),\ \hat{\bm\Sigma} \right]$, $j \in \{ 1, \dots, J \}$.
\item For each $(\bm\alpha^{(j)}, \bm\beta^{(j)})$, derive a sample $\theta^{(j)}$ of the HMM parameter of interest (i.e., transition probability or observation parameter), based on Equation \ref{eqn:linpred}.
\item Use quantiles of $\{  \theta^{(1)}, \theta^{(2)}, \dots, \theta^{(J)} \}$ as the bounds of a confidence interval for $\theta$ (e.g., the 2.5\% and 97.5\% quantiles for a 95\% confidence interval).
\end{enumerate}

Note that, even though the $\bm\lambda^{(j)}$ are not used in steps 2 and 3, it is important to sample from the joint distribution in step 1 to account for the uncertainty in the smoothness parameters. Choosing a large value for $J$ leads to a better approximation, but also to increased computational effort. In \pkg{hmmTMB}, $J = 1000$ is used as the default, but it can be changed by the user in plotting and prediction functions. 

In cases where $\theta$ is a function of covariates, it can be derived over a grid of covariate values in step 2, and the procedure then returns a pointwise confidence interval in step 3. Under the view that the smoothness penalty captures a prior distribution for $\bm\beta$, these are approximate credible intervals for penalised splines, which have the correct coverage probability ``across the function'' \citep[as defined by][Section 6.10]{wood2017}.

\section[hmmTMB overview]{\pkg{hmmTMB} overview}

The \pkg{hmmTMB} package is available on the Comprehensive \proglang{R} Archive Network (CRAN) at \url{https://CRAN.R-project.org/package=hmmTMB}, and its functionalities are described in several detailed vignettes, including example applications to financial and ecological data. Each function is also documented using \pkg{roxygen2} \citep{wickham2022}. Here, we provide an overview of the key features of the package, but we refer to the documentation and vignettes for more details. A simulation study is described in Appendix \ref{app:sim}, to investigate the performance of the method to estimate non-linear covariate effects and random effects.

\subsection{Installation}

The package can be installed using \code{install.packages("hmmTMB")}. It depends on the following \proglang{R} packages, all available from CRAN: \pkg{TMB} \citep{kristensen2016}, \pkg{mgcv} \citep{wood2017}, \pkg{RcppEigen} \citep{bates2013}, \pkg{R6} \citep{chang2021}, and \pkg{ggplot2} \citep{wickham2016}. For all code examples in this paper, we used \proglang{R} version 4.3.3, \pkg{TMB} version 1.9.17, \pkg{mgcv} version 1.9.1, \pkg{RcppEigen} version 0.3.4.0.2, \pkg{R6} version 2.6.1, and \pkg{ggplot2} version 3.5.1.

\subsection{Data requirements}

The data set for an HMM analysis needs to be provided as a \code{data.frame} object with one row for each observation time $t \in \{ 1, 2, \dots, n \}$. The data frame should include named columns for all variables that are included in the model, either as response variables or as covariates. It should also contain a column named ``\code{ID}'', to identify the time series in cases where multiple time series are analysed jointly. The likelihood of the full data set is computed as the product of the likelihoods of the individual time series (assuming independence). If \code{ID} is not provided, all observations are assumed to belong to the same time series. There is one other reserved column name: ``\code{state}'' should only be included when some states are known (Section \ref{sec:supervised}).

The package can accommodate missing values in the response variables, and these should be entered as \code{NA} in the data frame passed as input. However, model fitting does not allow for missing values in the covariates. If a covariate includes \code{NA}s, these are replaced by the last non-missing covariate value (or the \emph{next} non-missing value if there is no previous non-missing value). This is a crude solution, and each user should think about the best method to interpolate the covariate for their application. Variables included as random effects should be formatted as factors, as required by \pkg{mgcv}, rather than character strings or integers.

\subsection[R6 syntax]{\pkg{R6} syntax}

We use an object-oriented framework based on the package \pkg{R6} \citep{chang2021}. In this context, a class refers to a type of data structure, for which are defined some attributes (data and parameters) and some methods (functions specific to this class), and an object is an instance from a class. The general syntax is shown in the following code chunk.
\begin{knitrout}
\definecolor{shadecolor}{rgb}{0.969, 0.969, 0.969}\color{fgcolor}\begin{kframe}
\begin{alltt}
\hlcom{# Create object of class Class}
\hldef{object} \hlkwb{<-} \hldef{Class}\hlopt{$}\hlkwd{new}\hldef{()}
\hlcom{# Apply method() on object}
\hldef{object}\hlopt{$}\hlkwd{method}\hldef{()}
\end{alltt}
\end{kframe}
\end{knitrout}

In \pkg{hmmTMB}, there are three main classes, each representing a statistical model: \code{MarkovChain}, \code{Observation}, and \code{HMM} (described in more detail below). The general workflow is to first create an object (representing a model), and then apply methods to update its attributes (e.g., to fit the model). The object is always updated directly, and so results are always stored in the same object. In the rest of this paper, we refer to a particular method as \code{Class\$method()} (to make it clear which class it belongs to) even though, in practice, the method would be called on an object of that class.

One advantage of the object-oriented approach is that the hidden state model and observation model are created and stored separately in \pkg{hmmTMB}. As a result, one can for example simulate from a Markov chain, or plot state-dependent observation distributions, without the need to create a full HMM.

\subsection{Main classes}

We present the three main classes in \pkg{hmmTMB}, which are used to specify a hidden Markov model object.

\subsubsection[MarkovChain class]{\code{MarkovChain} class}

The \code{MarkovChain} class stores attributes related to the state process model, including the number of states, the transition probability matrices, and the formulas specifying the dependence of transition probabilities on covariates. The constructor \code{MarkovChain\$new()} takes the following arguments.
\begin{itemize}
\item \code{data}: data frame, required to define model matrices. This is needed even when no covariates are included, to determine the number of time points in the data.
\item \code{formula}: optional argument for model formulas. This can be either a single formula, which is then used for all transition probabilities, or a matrix of characters where each non-diagonal entry is the formula for a transition probability (and diagonal entries are set to \code{"."}).
\item \code{n\_states}: number of states $\geq 2$ (optional if \code{tpm} is provided).
\item \code{tpm}: transition probability matrix used to initialise the model. If not provided, the matrix is initialised with \code{0.9} along the diagonal and \code{0.1/(n\_states-1)} elsewhere.
\item \code{initial\_state}: character string indicating the model assumption used for the initial state distribution $\bm\delta^{(1)}$ of the Markov chain. Two important options are \code{"estimated"} (default), if $\bm\delta^{(1)}$ should be estimated, and \code{"stationary"}, if $\bm\delta^{(1)}$ should be fixed to the stationary distribution of the transition probability matrix $\bm\Gamma^{(1)}$.
\end{itemize}

\subsubsection[Observation class]{\code{Observation} class}

An object of the \code{Observation} class represents an observation model, defined by the list of observation variables and associated distributions, the state-dependent observation pararameters, and relevant formulas if observation parameters dependent on covariates. To create an \code{Observation} object, the constructor \code{Observation\$new()} requires the following arguments.
\begin{itemize}
\item \code{data}: data frame, containing at least the observation variables and covariates.
\item \code{dists}: named list of observation distributions for response variables.
\item \code{formulas}: optional nested list of formulas for the observation parameters. Each element corresponds to one of the observation variables, and it is itself a list, with one element for each parameter of the distribution. If this is not provided, no covariate dependence is assumed. 
\item \code{n\_states}: number of states $\geq 2$.
\item \code{par}: nested list of initial parameter values, with similar structure to \code{formulas}.
\end{itemize}

Families of distributions that are implemented for the observation model include distributions for continuous unbounded data (e.g., normal \code{"norm"}), continuous bounded data (e.g., exponential \code{"exp"}, gamma \code{"gamma"}, beta \code{"beta"}), discrete or binary data (e.g., Poisson \code{"pois"}, binomial \code{"binom"}), among others. The full list of distributions is given in Appendix \ref{app:dists}, and additional distributions will be added in the future. 

\subsubsection[HMM class]{\code{HMM} class}

The two main components of an \code{HMM} object are a \code{MarkovChain} object and an \code{Observation} object, which jointly describe the full specification of the model. These are the only two required arguments of the constructor \code{HMM\$new()}. \code{HMM} is the main class that users will interact with, as it includes methods to perform most steps of a typical HMM analysis: model fitting, state decoding, uncertainty quantification, model checking, and plotting.

\section{Statistical inference}

We summarise the main steps of a typical HMM analysis, and the key functions in the \pkg{hmmTMB} package.

\subsection{Model specification}

The state process model and the observation model are specified separately, as \code{MarkovChain} and \code{Observation} objects, respectively. An HMM is then created by combining them.

\subsubsection{Choice of number of states}

The choice of the number of states needs to be made prior to model fitting. There is no general statistical method to estimate the ``best'' number of states, and standard model selection criteria (e.g., AIC, BIC) often favour models with many states, even when they cannot be interpreted \citep{langrock2015}. \cite{pohle2017} suggest using domain expertise and model checking to choose the number of states for a given study.

\subsubsection{Formula syntax}

Model formulas can be defined when creating a \code{MarkovChain} or an \code{Observation} object, to allow for covariate effects. The formulas do not have a left-hand side, and the syntax for the right-hand side is borrowed from \pkg{mgcv}, as that package is used to create model matrices. Generally, base \proglang{R} formula syntax can be used for linear terms, e.g., \code{~x1 + x2 * x3} to include the linear effects of \code{x1}, \code{x2}, \code{x3}, and the interaction of \code{x2} and \code{x3}. Similarly, \code{poly()} can be used directly for polynomial terms.

For non-linear (``smooth'') terms, we use the function \code{s()} from \pkg{mgcv}, and we refer to its documentation for a detailed description. For example, we might use \code{~s(x1, k = 10, bs = "cs")} to model the non-linear relationship between a parameter and \code{x1}, where \code{k} and \code{bs} determine the dimension and type of basis used to define the spline. Some relevant choices of \code{bs} are \code{"cs"} (cubic spline), \code{"ts"} (thin plate regression spline), and \code{"cc"} (cyclical spline). All three are ``shrinkage'' bases: the spline is shrunk to zero when $\lambda \rightarrow \infty$, such that it can be effectively excluded from the model as part of smoothness estimation \citep[][Section 5.4.3]{wood2017}. The choice of \code{k} presents a trade-off between increased flexibility (large \code{k}) and reduced computational cost (small \code{k}). Past a certain point, increasing \code{k} will not change the model, as the smoothness of the spline is induced by the penalty. 

Independent normal random intercepts are \code{s(x, bs = "re")}, where \code{x} is the factor variable used as group. The \pkg{mgcv} syntax also allows for interactions between covariates, including interactions between continuous and categorical variables (\code{s(x, by = y, ...)} where \code{y} is a factor), and two-dimensional smooths with equal smoothness along each dimension (\code{s(x, by = y, ...)} where \code{y} is numeric). Non-isotropic smooth interactions with tensor products are not currently supported.

For formulas in the observation model, \pkg{hmmTMB} allows for special functions of the form \code{state1()}, \code{state2()}, etc., to include covariate effects only in certain states. For example, one might want to estimate the effect of some covariate on the mean of the observation distribution in state 1, but not in state 2, using something like \code{state1(s(x, k = 10, bs = "cs"))}. 

\subsection{Parameter estimation}
\label{sec:par}

The main method for parameter estimation is \code{HMM\$fit()}, which loads the (marginal) negative log-likelihood function from \pkg{TMB}, and optimises it numerically using the function \code{nlminb()}. Control settings such as \code{eval.max} and \code{iter.max} can optionally be passed to \code{HMM\$fit()} as additional arguments. Initial parameter values are required as a starting point for the optimiser, and their choice can be difficult for complex models; this is discussed in Section \ref{sec:initpar}.

\subsubsection{Accessing the parameter estimates}

After the model has been fitted by calling \code{HMM\$fit()}, all model parameters are automatically updated to their estimates in the \code{MarkovChain} and \code{Observation} components. There are multiple methods in \pkg{hmmTMB} to access the different model parameters:
\begin{itemize}
\item \code{HMM\$par()} returns estimates of the observation parameters $\bm\omega$ and transition probabilities $\bm\Gamma$. By default, it computes parameters for the first row of data when covariates are included, by this can be changed with the argument \code{t} (a vector of row indices for which the parameters should be computed).
\item \code{HMM\$coeff\_fe()} returns the fixed effect parameters $\bm\alpha$, such as intercepts and linear effects.
\item \code{HMM\$lambda()} returns the smoothness parameters $\bm\lambda$. Conversely, \code{HMM\$sd\_re()} returns the standard deviations of random effect terms, i.e., $1/\sqrt{\lambda_i}$.
\item \code{HMM\$coeff\_re()} returns predictions of the random effect parameters $\bm\beta$, e.g., levels of a random intercept, or basis coefficients for a spline. These are for example used to generate plots of the non-linear relationships.
\end{itemize}

\subsubsection{Uncertainty quantification and prediction}

The method \code{HMM\$predict()} can be used to predict the HMM parameters $\bm\omega$ and $\bm\Gamma$, including confidence intervals, possibly for user-defined covariate values. It takes the following arguments.
\begin{itemize}
\item \code{what}: the model component that should be predicted, either \code{"tpm"} (transition probabilities), \code{"delta"} (stationary distribution of state process, computed from transition probability matrix), or \code{"obspar"} (observation parameters).
\item \code{t}: time point, or vector of time points, for which the parameters should be predicted. This is only used if \code{newdata} is not specified, i.e., if the original data set is used.
\item \code{newdata}: data frame with covariate values for which the parameters should be predicted. If this is not provided, then the original data set is used.
\item \code{n\_post}: number of posterior samples to use to compute confidence intervals (as described in Section \ref{sec:implementation}). This defaults to 0, i.e., confidence intervals are not returned.
\end{itemize}

If confidence intervals are computed, the output is a list with three elements: \code{mean} (point estimate), \code{lcl} (lower bound), and \code{ucl} (upper bound). Each element is an array with one layer for each time point, and this output can for example be used to create plots of model parameters as functions of covariates, with confidence bands. 

Confidence intervals for the fixed effect parameters $\bm\alpha$ and smoothness parameters $\bm\lambda$ can be computed with \code{HMM\$confint()}, for example to quantify uncertainty on the linear effect of a covariate. These are derived by \code{TMB::sdreport()}, and the computational details are described in the documentation for that function.

\subsection{State decoding}

In many analyses, a key output is the classification of observations into states, i.e., an estimate of the hidden state $S_t$ for each $t \in \{ 1, 2, \dots, n \}$. This procedure is sometimes called state decoding \citep{zucchini2017}, and two different approaches are implemented in \pkg{hmmTMB}: global and local decoding. Given parameter estimates and the data, global decoding returns the most likely sequence of states. This is the sequence $(\hat{s}_1, \hat{s}_2, \dots, \hat{s}_n)$ which maximises the likelihood of the data, where $\hat{s}_t \in \{ 1, \dots, N \}$ is the estimated state for time $t$. The most common way to find this sequence is the Viterbi algorithm, and this is implemented in the method \code{HMM\$viterbi()} in \pkg{hmmTMB} \citep{viterbi1967}. In contrast, local decoding returns the probability of being in each state at each time point, $\Pr(S_t = j \vert Z_1, \dots, Z_n)$. The method \code{HMM\$state\_probs()} computes these probabilities based on the forward-backward algorithm \citep{zucchini2017}, and returns them as a matrix with one row for each time point and one column for each state. 

\subsection{Model checking}

Model checking is crucial to determine whether the assumptions of the model are appropriate. A popular approach to model checking for HMMs is the use of pseudo-residuals, which are constructed such that they are independent and follow a standard normal distribution if the model formulation is correct \citep{zucchini2017}. Similarly to linear regression residuals, deviations from normality indicate lack of fit, which can point to a failure of the observation distributions to model the data. In \pkg{hmmTMB}, pseudo-residuals can be computed with the method \code{HMM\$pseudores()}. This returns a matrix with one row for each response variable and one column for each observation. To assess goodness-of-fit for the $j$-th variable in the model, we can then extract the $j$-th row of the matrix, and use tools such as quantile-quantile plots (\code{qqnorm()}) and autocorrelation function plots (\code{acf()}) to investigate deviations from normality and independence, respectively.

Alternatively, we can use posterior predictive checks for model assessment. The general idea is to simulate from the fitted model, and compare the simulated and observed data, to identify features that are not captured by the model. The method \code{HMM\$check()} computes statistics, defined by the user through the argument \code{check\_fn}, for the observed data and for a large number of simulated data sets. If the observed statistic falls in the far tail of the simulated distribution, this suggests that the corresponding feature of the data-generating process was not captured well by the model. 

\subsection{Model visualisation}

In most cases, visualisation of the fitted model can greatly help with interpretation. Several methods are implemented in \pkg{hmmTMB} to plot the results of an analysis. We describe them briefly here, but they are presented in more detail in the example analysis of Section \ref{sec:example}, including figures showcasing their outputs. All plots are created with \pkg{ggplot2}, such that they can easily be edited by the user, e.g., to adjust captions, colour palettes, or axis scales.

\code{HMM\$plot()} is a flexible function to plot model parameters as functions of covariates, including confidence bands. The only two required arguments are ``\code{what}'' (which can be \code{"obspar"} for observation parameters, \code{"tpm"} for transition probabilities, or \code{"delta"} for stationary state probabilities), and ``\code{var}'' (name of covariate). \code{HMM\$plot\_ts()} creates one-dimensional or two-dimensional plots of data variables coloured by the most likely state sequence returned by the Viterbi algorithm. \code{HMM\$plot\_dist()} produces a graph of the state-dependent distributions, combined with a histogram of the observations. The state-dependent distributions are weighted by the proportion of time spent in each state in the Viterbi state sequence, so that the sum of the weighted distributions can be compared to the histogram.

\subsection{Numerical errors and starting parameters}
\label{sec:initpar}

Numerical errors can arise during model fitting if the optimiser fails to converge to the global maximum of the likelihood. For example, it might return an error if the log-likelihood is evaluated to be infinite, suggesting that the likelihood is very close to zero. In other cases, the package might fail to estimate uncertainty for the model parameters. Each situation is unique, and it is impossible to provide general guidelines to resolve such problems, but we would like to suggest two main approaches: simplifying the model formulation, and trying different sets of initial parameters. 

Complex model formulations require the estimation of a large number of parameters, and this can be a difficult task for the optimiser. In our experience, numerical errors are rare for 2-state covariate-free models, and this may therefore be a good place to start. Then, complexity may be increased incrementally, for example adding one covariate or one state at a time. Each time, the argument \code{init} of \code{HMM\$new()} can be used to initialise the parameters of the new (more complex) model to the estimated values from the previous (simpler) model. This approach reduces the risk of numerical errors, helps with choosing a parsimonious model formulation, and makes troubleshooting easier.

The performance of the optimiser depends on the choice of initial parameters, where starting values closer to the truth will generally perform better. It is therefore important to choose initial values that are plausible, based on inspection of the data, and we recommend trying several sets of starting values to ensure that the optimiser always converges to the same model. In \pkg{hmmTMB}, the method \code{Observation\$suggest\_initial()} can help with selecting initial parameters. This function uses the K-means algorithm, a simple clustering procedure which ignores the temporal structure of the data, to group observations into tentative ``states'' \citep{james2013}. It is then often straightforward to estimate the observation parameters from data within each group, and these are the suggested values. This is of course a crude approach but, in our experience, it usually performs well enough to circumvent numerical problems.

\subsection{Bayesian inference using Stan}
\label{sec:bayes}

The main workflow of \pkg{hmmTMB} presented in this section is based on the numerical optimisation of the likelihood function implemented in \code{HMM\$fit()}. As described in Section \ref{sec:implementation}, \pkg{TMB} marginalises the likelihood over random effects using the Laplace approximation, and it is the marginal likelihood that we optimise. 

An alternative approach is to use Markov chain Monte Carlo to sample over random effects (rather than integrate them out), and this can be done with \code{HMM\$fit\_stan()}. This method relies on the package \pkg{tmbstan}, which automatically uses a \pkg{TMB} model object to generate the equivalent \pkg{Stan} code \citep{monnahan2018, stan2022}. The model is then fitted using Hamiltonian Monte Carlo (HMC) methods, and posterior samples for all model parameters are returned. An example Bayesian analysis is included in Appendix \ref{app:activ}.

\subsubsection{Prior specification}

By default, improper flat priors are set for all model parameters, but these can be changed by the user. Specifically, normal priors can be specified for all fixed effect parameters $\bm\alpha$ and smoothness parameters $\bm\lambda$, using the method \code{HMM\$set\_priors()}. This function takes as input a list of matrices, where each matrix corresponds to a different model component (e.g., fixed effect parameters for hidden state model, fixed effect parameters for observation model). Each matrix has one row for each parameter of the corresponding model component, and two columns, for the means and standard deviations of the normal priors, respectively. A row set to \code{NA} corresponds to an improper flat prior. If there is demand, prior distributions could be extended to all observation distributions implemented in \pkg{hmmTMB} (listed in Appendix \ref{app:dists}).

\subsubsection{Model fitting}

The model can then be fitted using the method \code{HMM\$fit\_stan()}, which is a wrapper for \code{tmbstan::tmbstan()}. The main arguments required by this method are \code{chains}, the number of HMC chains to run, and \code{iter}, the number of HMC iterations in each chain. The choice of these settings will be highly context dependent, and we refer readers to the \pkg{Stan} documentation for more information. By default, \pkg{tmbstan} samples both fixed and random effects; alternatively, a hybrid approach can be used where fixed effects are sampled and random effect are integrated using the Laplace approximation, with the argument \code{laplace = TRUE}.

Once the algorithm has run, the output is an object of class \code{stanfit} (from the \pkg{rstan} package), and it can be accessed with \code{HMM\$out\_stan()}. This can either be transformed into a matrix with \code{as.matrix()} to inspect the posterior samples, or it can directly be passed to \pkg{rstan} functions for visualisation. In particular, \code{rstan::stan\_trace()} and \code{rstan::stan\_dens()} create trace and density plots of the posterior samples, respectively. Alternatively, \pkg{hmmTMB} automatically computes posterior samples for the HMM parameters directly (rather than linear predictor parameters), and these can be accessed using \code{HMM\$iters()}. This method returns a matrix with one row for each HMC iteration, and one column for each HMM parameter (observation parameters and transition probabilities). These transformed posterior draws are computed for the first time point of the data set, and are of limited utility in models where parameters are time-varying, i.e., if covariates are included.

\subsubsection{Comparison to maximum likelihood estimation}

The choice of using the Bayesian or maximum likelihood approach will depend on each user's preference and analysis. For example, the Bayesian method has the appeal of including prior information when it is available, and allowing for full posterior inference, but this comes at a computational cost which might make the approach infeasible for large data sets or complex models.

In \pkg{hmmTMB}, a key difference between the two approaches is the way they deal with random effects. The maximum (marginal) likelihood fitting procedure is based on the Laplace approximation to integrate over random effects. This makes the assumption that the likelihood can be approximated reasonably well by a normal density function. The HMM likelihood is often a complex function with multiple (local) modes \citep{zucchini2017}, which suggests that the Laplace approximation might not be appropriate in some cases. Research is needed to investigate this problem, and we hope that \pkg{hmmTMB} can greatly help with this, as it allows for both approaches to be applied to the same model, and outputs to be compared.

\section{Other features}

\subsection{Simulation}

The method \code{HMM\$simulate()} simulates observations from an HMM, either using the estimated parameters if the model has been fitted, or using the initial parameters otherwise (e.g., to simulate from a user-defined HMM). The state sequence is first simulated from the Markov chain model, and the observations are then generated based on the simulated states. In models with covariates, the function requires a data frame with one column for each covariate, to be passed as the argument \code{data}. This function is used in \code{HMM\$check()} to compare user-defined features of the observed data to simulated data, for model assessment.

\subsection{Supervised learning}
\label{sec:supervised}

In most applications, HMMs are used in an unsupervised setting: the states are not known prior to the analysis, and they are entirely data-driven. However, in some cases, information might be available about the states before the analysis. For example, the analyst might be able to classify observations into pre-defined states manually for a subset of the data, but not for the full data set because of time constraints. Then, the aim of the analysis is two-fold: characterising the pre-defined states with a statistical model, and classifying the full data set. This approach is sometimes called ``semi-supervised'' \citep{leos2017}.

Supervised (or semi-supervised) learning can be implemented in \pkg{hmmTMB}, by including a column named \code{state} in the input data frame. In a $K$-state model, this column should contain numbers between 1 and $K$ for time points where the state is known, and \code{NA} elsewhere. It is detected by the package, and used to fix the known states. 

\subsection{Fixed parameters}
\label{sec:fixed}

In some analyses, it might be useful to fix a parameter to a given value, rather than estimate it. This might for example be the case if a transition is impossible, i.e., the corresponding transition probability should be fixed to zero. Alternatively, we might sometimes want to constrain several parameters in the model to be estimated to a shared value. Such model formulations can be implemented in \pkg{hmmTMB} using the \code{fixpar} argument of \code{HMM\$new()}. This is based on the parameter mapping functionality offered by \pkg{TMB}, with the argument \code{map} of \code{TMB::MakeADFun()}, and we refer the reader to the \pkg{TMB} documentation for details.

Specifically, \code{fixpar} is defined as a named list with the following optional elements: \code{obs} (observation parameters), \code{hid} (transition probabilities), \code{lambda\_obs} (smoothness parameters for observation model), \code{lambda\_hid} (smoothness parameters for state model), and \code{delta0} (initial distribution of state process). Each of these entries should be a named vector, in a format similar to that expected for the argument \code{map} of \code{TMB::MakeADFun()}. Each element of the vector should be named after a parameter, and its value should be \code{NA} for a fixed parameter, to indicate that it should be fixed to its initial value. If several parameters have a common value, the corresponding elements in \code{fixpar} should be set to some shared integer. The relevant parameter names can be found using \code{HMM\$coeff\_list()}.

\subsection{General dependence structures}
\label{sec:general_dependence}

The flexible parameter constraints presented in Section \ref{sec:fixed} make it possible to implement HMMs with non-standard dependence structures in \pkg{hmmTMB}. These models relax either the Markov assumption of the state process, or the assumption that successive observations are independent conditional on the state. These extensions often come with greatly increased computational cost, because they suffer from the curse of dimensionality (the number of parameters increases quickly with model complexity). Parameters from these more complex methods are also often difficult to interpret, and so we recommend parsimony when choosing a model formulation for a given analysis.

\paragraph{Hidden semi-Markov models} The dwell times of a Markov chain (sometimes called sojourn times) are the time intervals between state transitions. As a consequence of the Markov assumption (Equation \ref{eqn:markov}), dwell times in state $j$ follow a geometric distribution with parameter $1 - \gamma_{jj}$. Semi-Markov models relax the Markov assumption, and increase flexibility by explicitly modelling the dwell time distribution \citep[][Chapter 12]{zucchini2017}. It has been shown that a hidden semi-Markov model can be approximated by a standard HMM with expanded state space \citep{langrock2011}. Briefly, this requires representing each state of the semi-Markov model with several states in a Markov chain, and constraining the transition probabilities to control the time spent in each composite state. The approach of \cite{langrock2011} can be implemented in \pkg{hmmTMB} by forcing the relevant transition probabilities to be fixed to zero. This method can in principle be used to estimate any dwell time distribution with arbitrary precision but, in practice, the memory requirements might be prohibitive.

\paragraph{Higher-order hidden Markov models} Another approach to relax the Markov assumption is to allow the state $S_t$ to depend not only on $S_{t-1}$, but also on $S_{t-2}$ (and possibly $S_{t-3}, S_{t-4}$, etc.). The resulting model is called a higher-order Markov chain, which can incorporate more ``memory'' about the history of the state process than the first-order model \citep[][Section 10.3]{zucchini2017}. A higher-order Markov chain $(\tilde{S}_t)$ can be viewed as a first-order Markov chain $(S_t)$ by defining $S_t = (\tilde{S}_t, \tilde{S}_{t-1}, \dots, \tilde{S}_{t-k+1})$. For example, if $(\tilde{S}_t)$ is a 2-state second-order Markov chain, then $(S_t)$ is a 4-state first-order Markov chain, with state space $\{ (1, 1), (1, 2), (2, 1), (2, 2) \}$. Some transition probabilities in this 4-state model must be fixed to zero; for example, it is impossible to transition from $S_t = (\tilde{S}_{t}, \tilde{S}_{t-1}) = (1, 1)$ to $S_{t+1} = (\tilde{S}_{t+1}, \tilde{S}_{t}) = (1, 2)$.

\paragraph{Autoregressive hidden Markov models} The assumption that successive observations of an HMM are conditionally independent given the state process can be relaxed, for example to capture strong autocorrelation in the observation process. A practical approach is to include the observation $\bm{Z}_{t-1}$ (and possibly $\bm{Z}_{t-2}$ and so on) as a covariate on the parameters of the distribution of $\bm{Z}_t$. For example, a Markov-switching first-order autoregressive model could be written as $Z_t \mid \{ S_t = j \} \sim N(\alpha_j Z_{t-1},\ \sigma_j^2 )$. The implementation in \pkg{hmmTMB} would require creating an observation model that includes the linear effect of a lagged version of the observed variable on the mean parameter (and no intercept). The slope of this effect is $\alpha_j$. A similar approach can be used to implement Markov-switching Gaussian random walks, correlated random walks, and higher-order autoregressive models. 

\paragraph{Coupled hidden Markov models} A coupled HMM models two observed variables $Z_t^{(1)}$ and $Z_t^{(2)}$ as being driven by two separate, but dependent, state processes $(S_t^{(1)})$ and $(S_t^{(2)})$, respectively \citep{pohle2021}. Several possible assumptions can be made about the dependence of the state processes, and here we focus on the Cartesian product model of \cite{pohle2021}, which can be implemented in \pkg{hmmTMB}. The Cartesian product model is based on the process $(S_t)$ where $S_t = (S_t^{(1)}, S_t^{(2)})$. For example, if $(S_t^{(1)})$ and $(S_t^{(2)})$ are 2-state Markov chains, then $(S_t)$ is a 4-state Markov chain with state space $\{ (1, 1), (1, 2), (2, 1), (2, 2) \}$. The coupled model can therefore be implemented as a 4-state HMM with some constraints on the observation parameters: the distribution parameters of $Z_t^{(1)}$ are the same in states $S_t = 1$ and $S_t = 2$ (because both imply $S_t^{(1)} = 1$), and so on.

\subsection{Accessing TMB outputs}

Users familiar with \pkg{TMB} may be interested to directly access and manipulate its outputs, for example to evaluate the negative log-likelihood of the model, or to find the covariance matrix of the parameters. The method \code{HMM\$tmb\_obj()} returns the object created by \code{TMB::MakeADFun()}. This is a list which includes \code{fn} (a function object for negative log-likelihood), and \code{gr} (function object for gradient of negative log-likelihood). After likelihood optimisation, \pkg{hmmTMB} uses the function \code{TMB::sdreport()}, which computes covariance (or precision) matrices for all model parameters, used for uncertainty quantification. The output of \code{TMB::sdreport()} can be accessed using the method \code{HMM\$tmb\_rep()}.

\section{Comparison to other packages}

There are many software packages for HMMs, which greatly differ in the model formulations that they include, and their focus on particular data types. Although a comprehensive review is infeasible, we contrast the features of some of the most popular and flexible packages to \pkg{hmmTMB}, to place it within its broader context.

In \proglang{R}, two popular packages are \pkg{depmixS4} and \pkg{msm}, which both provide great flexibility for model formulation \citep{visser2010, jackson2011}. In particular, similarly to \pkg{hmmTMB}, they can accommodate multivariate data, (linear) covariate effects on transition probabilities, missing data, multiple time series, and parameter constraints. Although \pkg{depmixS4} includes only a few observation distributions by default, it allows for user-defined distributions. It might therefore be a good alternative in cases where a distribution which is not included in \pkg{hmmTMB} is needed. \pkg{depmixS4} also allows for more complex parameter constraints than \pkg{hmmTMB}, for example inequalities between model parameters. \pkg{msm} has a focus on continuous-time HMMs, which are for example widely used in medicine. Although \pkg{hmmTMB} does not currently support continuous-time HMMs, we plan to include this functionality in a future release.

The \proglang{R} packages \pkg{moveHMM} and \pkg{momentuHMM} are also widely used to apply HMMs, with a focus on animal tracking studies in ecology \citep{michelot2016, mcclintock2018}. \pkg{moveHMM} is specialised for the analysis of two-dimensional animal movement data, making it easier to use than alternatives, but limiting its flexibility in other analyses. \pkg{momentuHMM} extends \pkg{moveHMM} to more general model formulations, and it allows for multivariate data, a wide range of observation distributions, parameter constraints, and (linear) covariate effects on all model parameters. In addition, it allows for discrete random effects, as described for example by \cite{maruotti2009} and \cite{mckellar2015}. In this approach, a parameter may take several discrete values, and each individual or group is allocated to one of the values. This stands in contrast with the continuous random effects implemented in \pkg{hmmTMB} (Section \ref{sec:model}).

In Python, the main HMM package is \pkg{hmmlearn} \citep{lebedev2022}. It includes three families of observation distributions (normal, normal mixture, and multinomial), and requires user-defined functions for other distributions. It can be used on multiple time series, and can include parameter constraints; however, it does not allow for covariate effects on the HMM parameters.

Overall, \pkg{hmmTMB} is one of the fastest and most flexible available packages for HMMs. Its most unique feature is the ability to implement very general non-parametric and hierarchical models for the transition probabilities and the observation parameters (Section \ref{sec:model}). This includes splines and their interactions, (continuous) random intercepts and random slopes, and the large family of hierarchical generalised additive models \citep{pedersen2019}. The interface with \pkg{Stan} will also greatly increase the accessibility of Bayesian inference methods for HMMs \citep[although see][]{damiano2023}. Finally, the package includes a few smaller features which we expect to be useful in many HMM analyses, such as model checking using posterior predictive simulations, and customisable visualisations based on \pkg{ggplot2}.

\section{Reproducible example}
\label{sec:example}

We illustrate the main features of \pkg{hmmTMB} through an example analysis of animal tracking data, including non-linear covariate effects and random effects in the state process model. Two additional examples with code are available in the supplementary material: analysis of human activity data with cyclical covariate effect in a Bayesian framework (Appendix \ref{app:activ}), and analysis of energy prices with regime-switching generalised additive model (Appendix \ref{app:energy}).

\subsection{Petrel movement analysis}

We considered time-indexed GPS locations from 10 Antarctic petrels, described by \cite{descamps2016paper} and available through the Movebank data repository \citep{descamps2016data}. HMMs are commonly used on such data to describe the animals' movement in terms of a few states, often interpreted as behavioural states \citep{patterson2009}. The variables of interest are the step lengths (distances between successive locations) and turning angles (angles between successive segments of the track), and we derived them using the \proglang{R} package \pkg{moveHMM} \citep{michelot2016}. These two variables are often informative about animal behaviour, because they capture both the speed and tortuosity of movement. In addition to step length and turning angle, we derived distances between each bird's locations and its ``central location'', taken to be the first location of its track, and interpreted as a colony. We were interested in modelling behavioural states of petrels, and understanding how their dynamics are driven by the distance to central location. Observations were obtained at regular 30-min intervals, and covered between one and three weeks for the different birds, resulting in 6296 observations for each variable. The code used to preprocess the Movebank data is given in Appendix \ref{app:petrel}.

\begin{knitrout}
\definecolor{shadecolor}{rgb}{0.969, 0.969, 0.969}\color{fgcolor}\begin{kframe}
\begin{alltt}
\hlcom{# Load package and data}
\hlkwd{library}\hldef{(hmmTMB)}
\hldef{data} \hlkwb{<-} \hlkwd{read.csv}\hldef{(}\hlsng{"petrels.csv"}\hldef{)}
\hlcom{# Transform ID to factor for random effects}
\hldef{data}\hlopt{$}\hldef{ID} \hlkwb{<-} \hlkwd{factor}\hldef{(data}\hlopt{$}\hldef{ID)}
\hldef{data}\hlopt{$}\hldef{time} \hlkwb{<-} \hlkwd{as.POSIXct}\hldef{(data}\hlopt{$}\hldef{time)}

\hlkwd{head}\hldef{(data)}
\end{alltt}
\begin{verbatim}
     ID                time  lon   lat   d2c  step    angle
1 PET-A 2011-12-13 00:22:57 7.23 -69.9  0.00  8.41       NA
2 PET-A 2011-12-13 00:52:57 7.26 -69.8  8.41 12.36 -0.04390
3 PET-A 2011-12-13 01:22:57 7.31 -69.7 20.76 13.44 -0.06130
4 PET-A 2011-12-13 01:52:57 7.39 -69.6 34.17 13.04 -0.07485
5 PET-A 2011-12-13 02:22:57 7.49 -69.5 47.14 13.07  0.00498
6 PET-A 2011-12-13 02:52:57 7.59 -69.4 60.17 12.68  0.03462
\end{verbatim}
\end{kframe}
\end{knitrout}

We modelled step length $L_t > 0$ with a gamma distribution, parameterised in terms of its state-dependent mean $\omega_{1j} > 0$ and standard deviation $\omega_{2j} > 0$, and turning angle $\varphi_t \in (-\pi, \pi]$ with a wrapped Cauchy distribution with mean zero, parameterised in terms of a state-dependent concentration parameter $\omega_{3j} \in [0, 1]$:
\begin{equation*}
  \begin{cases}
    L_t \mid \{ S_t = j \} \sim \text{gamma}(\omega_{1j}, \omega_{2j}) \\
    \varphi_t \mid \{ S_t = j \} \sim \text{wrapped Cauchy}(0, \omega_{3j})
  \end{cases}
\end{equation*}
The mean turning angle in each state is fixed to zero (rather than treated as an unknown parameter) because the petrels' movement has strong directional persistence, as is common in high-resolution tracking data.

We used a 3-state model, with the expectation that state 1 would capture slow movement, state 3 would capture fast movement, and state 2 would be intermediate. In the state process model, we included a random intercept for the individual $I_t \in \{ 1, 2, \dots, 10 \}$, to capture inter-individual variability \citep{mcclintock2021}. We also included a non-linear effect of the distance to centre $d_t > 0$ on the transition probabilities. We constrained the transition probabilities between states 1 and 3 ($\gamma_{13}$ and $\gamma_{31}$) to be zero, under the assumption that animals need to transition through the intermediate state 2. Finally, the hidden state model can be summarised with the following expressions of the linear predictor,
\begin{equation*}
  \eta_{ij}^{(t)} =
  \begin{cases}
    0  & \text{if } i = j\\
    - \infty & \text{if } (i, j) = (1, 3) \text{ or } (3, 1)\\
    \alpha_{ij} + \beta_{ij}^{(I_t)} + f_{ij}(d_t)  & \text{otherwise},
  \end{cases}
\end{equation*}
where $\beta_{ij}^{(k)} \sim N(0, \sigma_{ij}^2)$ is the random intercept for track $k$, and where $f_{ij}$ is a smooth function, modelled with a spline. 

\subsection{Model specification}

\subsubsection{Observation model}

Creating an \code{Observation} object requires defining a list of observation distributions; we use the \code{"gamma2"} distribution for step length (gamma distribution parameterised by mean and standard deviation), and the \code{"wrpcauchy"} distribution for turning angle (wrapped Cauchy distribution). We also need to define a list of initial parameter values for the optimiser; here, we choose them based on data visualisation, but we could also update them based on \code{Observation\$suggest\_initial()}.

\begin{knitrout}
\definecolor{shadecolor}{rgb}{0.969, 0.969, 0.969}\color{fgcolor}\begin{kframe}
\begin{alltt}
\hlcom{# Initial parameters}
\hldef{step_mean0} \hlkwb{<-} \hlkwd{c}\hldef{(}\hlnum{1}\hldef{,} \hlnum{6}\hldef{,} \hlnum{20}\hldef{)}
\hldef{step_sd0} \hlkwb{<-} \hlkwd{c}\hldef{(}\hlnum{1}\hldef{,} \hlnum{5}\hldef{,} \hlnum{10}\hldef{)}
\hldef{angle_mean0} \hlkwb{<-} \hlkwd{c}\hldef{(}\hlnum{0}\hldef{,} \hlnum{0}\hldef{,} \hlnum{0}\hldef{)}
\hldef{angle_rho0} \hlkwb{<-} \hlkwd{c}\hldef{(}\hlnum{0.8}\hldef{,} \hlnum{0.8}\hldef{,} \hlnum{0.9}\hldef{)}
\hldef{par0} \hlkwb{<-} \hlkwd{list}\hldef{(}\hlkwc{step} \hldef{=} \hlkwd{list}\hldef{(}\hlkwc{mean} \hldef{= step_mean0,} \hlkwc{sd} \hldef{= step_sd0),}
             \hlkwc{angle} \hldef{=} \hlkwd{list}\hldef{(}\hlkwc{mu} \hldef{= angle_mean0,} \hlkwc{rho} \hldef{= angle_rho0))}

\hlcom{# Observation distributions}
\hldef{dists} \hlkwb{<-} \hlkwd{list}\hldef{(}\hlkwc{step} \hldef{=} \hlsng{"gamma2"}\hldef{,} \hlkwc{angle} \hldef{=} \hlsng{"wrpcauchy"}\hldef{)}

\hlcom{# Create Observation object}
\hldef{obs} \hlkwb{<-} \hldef{Observation}\hlopt{$}\hlkwd{new}\hldef{(}\hlkwc{data} \hldef{= data,}
                       \hlkwc{dists} \hldef{= dists,}
                       \hlkwc{n_states} \hldef{=} \hlnum{3}\hldef{,}
                       \hlkwc{par} \hldef{= par0)}
\end{alltt}
\end{kframe}
\end{knitrout}

\subsubsection{Hidden state model}

If all transition probabilities had the same model formula (i.e., if they all depended on the same covariates), we could simply pass this formula to \code{MarkovChain\$new()}. However, in this example, the transition probabilities $\gamma_{13}$ and $\gamma_{31}$ do not depend on covariates, as they are fixed to zero. So, we need to create a matrix to specify the structure of the model, where each element is the formula (as character string) for the corresponding transition probability. The diagonal elements are used as references for each row, so their formulas are set to \code{"."}. We create a matrix of initial values for the transition probabilities, where $\gamma_{13}$ and $\gamma_{31}$ are set to zero. The other elements are used to initialise the intercepts for the corresponding transition probabilities.

\begin{knitrout}
\definecolor{shadecolor}{rgb}{0.969, 0.969, 0.969}\color{fgcolor}\begin{kframe}
\begin{alltt}
\hlcom{# Model formulas}
\hldef{f} \hlkwb{<-} \hlsng{"~ s(ID, bs = 're') + s(d2c, k = 10, bs = 'cs')"}
\hldef{tpm_structure} \hlkwb{<-} \hlkwd{matrix}\hldef{(}\hlkwd{c}\hldef{(}\hlsng{"."}\hldef{,   f,} \hlsng{"~1"}\hldef{,}
                          \hldef{f,}   \hlsng{"."}\hldef{,    f,}
                          \hlsng{"~1"}\hldef{,  f,}  \hlsng{"."}\hldef{),}
                        \hlkwc{ncol} \hldef{=} \hlnum{3}\hldef{,} \hlkwc{byrow} \hldef{=} \hlnum{TRUE}\hldef{)}

\hlcom{# Initial transition probabilities}
\hldef{tpm0} \hlkwb{<-} \hlkwd{matrix}\hldef{(}\hlkwd{c}\hldef{(}\hlnum{0.9}\hldef{,} \hlnum{0.1}\hldef{,} \hlnum{0}\hldef{,}
                 \hlnum{0.1}\hldef{,} \hlnum{0.8}\hldef{,} \hlnum{0.1}\hldef{,}
                 \hlnum{0}\hldef{,} \hlnum{0.1}\hldef{,} \hlnum{0.9}\hldef{),}
               \hlkwc{ncol} \hldef{=} \hlnum{3}\hldef{,} \hlkwc{byrow} \hldef{=} \hlnum{TRUE}\hldef{)}

\hlcom{# Create MarkovChain object}
\hldef{hid} \hlkwb{<-} \hldef{MarkovChain}\hlopt{$}\hlkwd{new}\hldef{(}\hlkwc{n_states} \hldef{=} \hlnum{3}\hldef{,}
                       \hlkwc{formula} \hldef{= tpm_structure,}
                       \hlkwc{data} \hldef{= data,}
                       \hlkwc{tpm} \hldef{= tpm0,}
                       \hlkwc{initial_state} \hldef{=} \hlsng{"stationary"}\hldef{)}
\end{alltt}
\end{kframe}
\end{knitrout}

We used the option \code{initial\_state = "stationary"} to fix the initial distribution $\bm\delta^{(1)}$ of the state process to the stationary distribution of $\bm\Gamma^{(1)}$ for each time series. We do this for computational convenience to circumvent common identifiability problems for the estimation of the initial distribution. However, note that the Markov process is not stationary due to effects of time-varying covariates, and so this does not define a stationary HMM \citep{zucchini2017}. Another approach would be to set \code{initial\_state} to a vector of integers, which would be used as the fixed initial state for each time series, if the initial states were known.

\subsubsection{Hidden Markov model}

We need to define two different parameter constraints: the means of the turning angle distributions are fixed to zero, and the transition probabilities between states 1 and 3 are fixed to zero. We have initialised the parameters to these values, and we now need to let the package know that they should not be estimated (i.e., kept fixed at the initial values), using the \code{fixpar} argument of \code{HMM\$new()}. We do this as explained in Section \ref{sec:fixed}, and we can then create the \code{HMM} object.

\begin{knitrout}
\definecolor{shadecolor}{rgb}{0.969, 0.969, 0.969}\color{fgcolor}\begin{kframe}
\begin{alltt}
\hlcom{# List of fixed parameters}
\hldef{fixpar} \hlkwb{<-} \hlkwd{list}\hldef{(}\hlkwc{obs} \hldef{=} \hlkwd{c}\hldef{(}\hlsng{"angle.mu.state1.(Intercept)"} \hldef{=} \hlnum{NA}\hldef{,}
                       \hlsng{"angle.mu.state2.(Intercept)"} \hldef{=} \hlnum{NA}\hldef{,}
                       \hlsng{"angle.mu.state3.(Intercept)"} \hldef{=} \hlnum{NA}\hldef{),}
               \hlkwc{hid} \hldef{=} \hlkwd{c}\hldef{(}\hlsng{"S1>S3.(Intercept)"} \hldef{=} \hlnum{NA}\hldef{,}
                       \hlsng{"S3>S1.(Intercept)"} \hldef{=} \hlnum{NA}\hldef{))}

\hlcom{# Create HMM object}
\hldef{hmm} \hlkwb{<-} \hldef{HMM}\hlopt{$}\hlkwd{new}\hldef{(}\hlkwc{obs} \hldef{= obs,} \hlkwc{hid} \hldef{= hid,} \hlkwc{fixpar} \hldef{= fixpar)}
\end{alltt}
\end{kframe}
\end{knitrout}

We can now fit the model with \code{HMM\$fit()}. This takes about 1 min on a desktop computer with a 13-generation Intel i9-13900 CPU.
\begin{knitrout}
\definecolor{shadecolor}{rgb}{0.969, 0.969, 0.969}\color{fgcolor}\begin{kframe}
\begin{alltt}
\hldef{hmm}\hlopt{$}\hlkwd{fit}\hldef{(}\hlkwc{silent} \hldef{=} \hlnum{TRUE}\hldef{)}
\end{alltt}
\end{kframe}
\end{knitrout}

\subsection{Results}

We can get estimates of all model parameters using the methods described in Section \ref{sec:par}. Here, we showcase plotting methods from the package, which can greatly help with model interpretation.

\subsubsection{Interpreting the states}

The most likely state sequence is computed by \code{HMM\$viterbi()}. It is often used to create a plot of the data, coloured by state, and this is implemented in \code{HMM\$plot\_ts()} (Figure \ref{fig:map}).

\begin{knitrout}
\definecolor{shadecolor}{rgb}{0.969, 0.969, 0.969}\color{fgcolor}\begin{kframe}
\begin{alltt}
\hldef{hmm}\hlopt{$}\hlkwd{plot_ts}\hldef{(}\hlsng{"lon"}\hldef{,} \hlsng{"lat"}\hldef{)} \hlopt{+}
  \hlkwd{coord_map}\hldef{(}\hlsng{"mercator"}\hldef{)} \hlopt{+}
  \hlkwd{geom_point}\hldef{(}\hlkwc{size} \hldef{=} \hlnum{0.3}\hldef{)} \hlopt{+}
    \hlkwd{labs}\hldef{(}\hlkwc{x} \hldef{=} \hlsng{"longitude"}\hldef{,} \hlkwc{y} \hldef{=} \hlsng{"latitude"}\hldef{)}
\end{alltt}
\end{kframe}\begin{figure}

{\centering \includegraphics[width=0.7\linewidth]{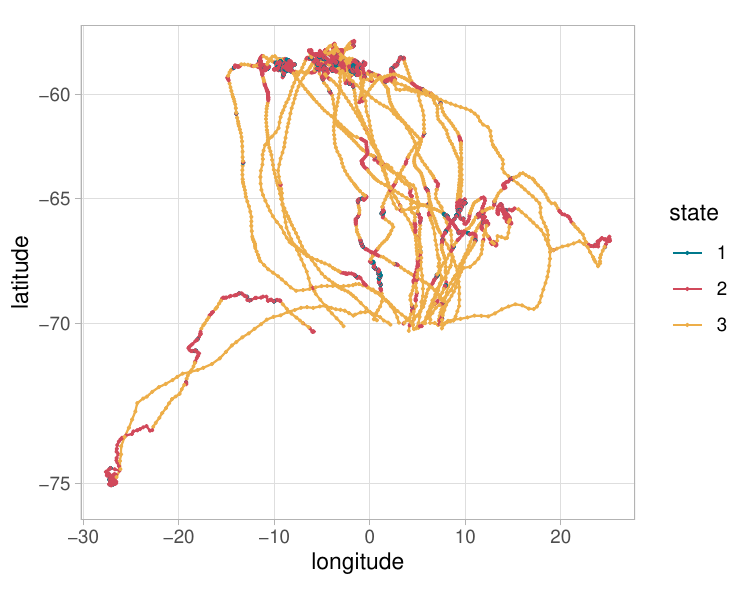} 

}

\caption[Map of petrel tracks, coloured by most likely state sequence]{Map of petrel tracks, coloured by most likely state sequence.}\label{fig:map}
\end{figure}

\end{knitrout}

Another useful output is plots of the estimated state-dependent density functions for step length and turning angle. These plots are generated by the method \code{HMM\$plot\_dist()}, and shown in Figure \ref{fig:dists}. 

\begin{knitrout}
\definecolor{shadecolor}{rgb}{0.969, 0.969, 0.969}\color{fgcolor}\begin{kframe}
\begin{alltt}
\hldef{hmm}\hlopt{$}\hlkwd{plot_dist}\hldef{(}\hlsng{"step"}\hldef{)} \hlopt{+}
    \hlkwd{coord_cartesian}\hldef{(}\hlkwc{ylim} \hldef{=} \hlkwd{c}\hldef{(}\hlnum{0}\hldef{,} \hlnum{0.25}\hldef{))} \hlopt{+}
    \hlkwd{theme}\hldef{(}\hlkwc{legend.position} \hldef{=} \hlkwd{c}\hldef{(}\hlnum{0.8}\hldef{,} \hlnum{0.8}\hldef{))} \hlopt{+}
    \hlkwd{labs}\hldef{(}\hlkwc{x} \hldef{=} \hlsng{"step (km)"}\hldef{)}

\hldef{hmm}\hlopt{$}\hlkwd{plot_dist}\hldef{(}\hlsng{"angle"}\hldef{)} \hlopt{+}
    \hlkwd{theme}\hldef{(}\hlkwc{legend.position} \hldef{=} \hlsng{"none"}\hldef{)} \hlopt{+}
    \hlkwd{scale_x_continuous}\hldef{(}\hlkwc{breaks} \hldef{=} \hlkwd{seq}\hldef{(}\hlopt{-}\hldef{pi, pi,} \hlkwc{by} \hldef{= pi}\hlopt{/}\hlnum{2}\hldef{),}
                       \hlkwc{labels} \hldef{=} \hlkwd{expression}\hldef{(}\hlopt{-}\hldef{pi,} \hlopt{-}\hldef{pi}\hlopt{/}\hlnum{2}\hldef{,} \hlnum{0}\hldef{, pi}\hlopt{/}\hlnum{2}\hldef{, pi))}
\end{alltt}
\end{kframe}\begin{figure}
\includegraphics[width=0.49\linewidth]{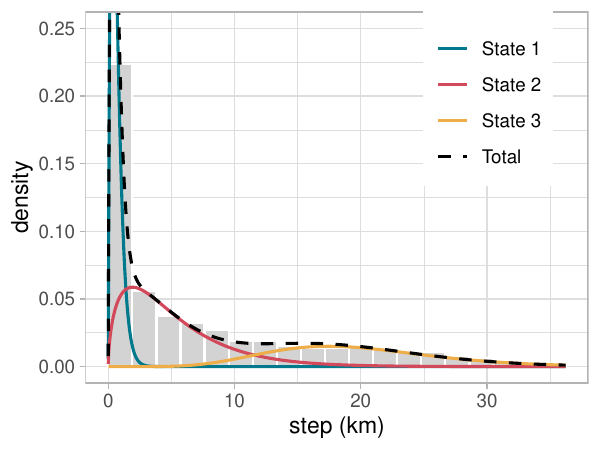} 
\includegraphics[width=0.49\linewidth]{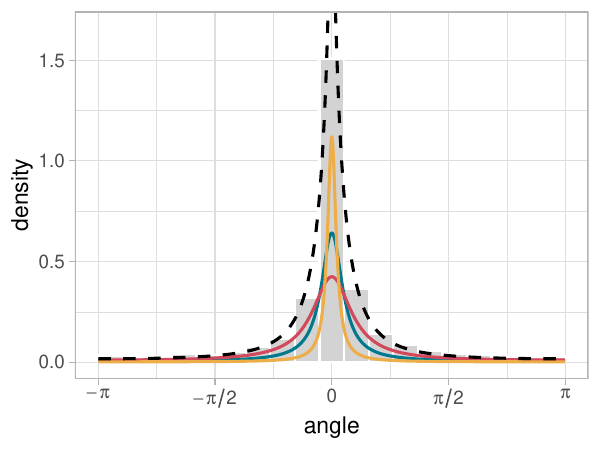} \caption[Estimated distributions of step length (left) and turning angle (right) in the petrel analysis]{Estimated distributions of step length (left) and turning angle (right) in the petrel analysis. The densities are weighted by the proportion of the states in the Viterbi sequence.}\label{fig:dists}
\end{figure}

\end{knitrout}

Both Figures \ref{fig:map} and \ref{fig:dists} suggest that the main difference between the states is in the speed of movement: state 1 captured slow movement, state 3 is fast movement, and state 2 is intermediate. All three states have high turning angle concentrations, i.e., strong directional persistence. In the following, we assume that these states correspond to three behaviours, ``foraging'' (state 1), ``exploring'' (state 2), and ``travelling'' (state 3), but we recognise that this likely ignores other important behaviours (e.g., resting on water).

\subsubsection{Effect of distance to centre}

The main aim of this analysis was to investigate drivers of behavioural switching in petrels. We can use the method \code{HMM\$plot()} to plot the transition probabilities, or the stationary state probabilities, as functions of covariates. 

\begin{knitrout}
\definecolor{shadecolor}{rgb}{0.969, 0.969, 0.969}\color{fgcolor}\begin{kframe}
\begin{alltt}
\hlcom{# Transition prob Pr(3 -> 2)}
\hldef{hmm}\hlopt{$}\hlkwd{plot}\hldef{(}\hlkwc{what} \hldef{=} \hlsng{"tpm"}\hldef{,} \hlkwc{var} \hldef{=} \hlsng{"d2c"}\hldef{,} \hlkwc{i} \hldef{=} \hlnum{3}\hldef{,} \hlkwc{j} \hldef{=} \hlnum{2}\hldef{)} \hlopt{+}
    \hlkwd{labs}\hldef{(}\hlkwc{x} \hldef{=} \hlsng{"distance to centre (km)"}\hldef{)}

\hlcom{# Stationary state probabilities}
\hldef{hmm}\hlopt{$}\hlkwd{plot}\hldef{(}\hlkwc{what} \hldef{=} \hlsng{"delta"}\hldef{,} \hlkwc{var} \hldef{=} \hlsng{"d2c"}\hldef{)} \hlopt{+}
    \hlkwd{theme}\hldef{(}\hlkwc{legend.position} \hldef{=} \hlsng{"top"}\hldef{,} \hlkwc{legend.margin} \hldef{=} \hlkwd{margin}\hldef{(}\hlkwd{c}\hldef{(}\hlnum{0}\hldef{,} \hlnum{0}\hldef{,} \hlopt{-}\hlnum{10}\hldef{,} \hlnum{0}\hldef{)))} \hlopt{+}
    \hlkwd{labs}\hldef{(}\hlkwc{title} \hldef{=} \hlkwa{NULL}\hldef{,} \hlkwc{x} \hldef{=} \hlsng{"distance to centre (km)"}\hldef{)}
\end{alltt}
\end{kframe}\begin{figure}
\includegraphics[width=0.49\linewidth]{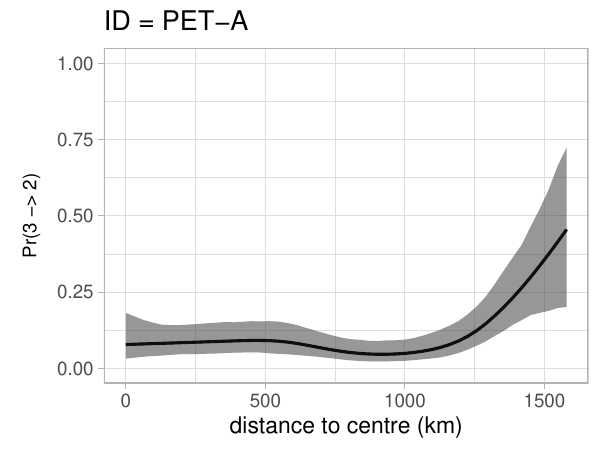} 
\includegraphics[width=0.49\linewidth]{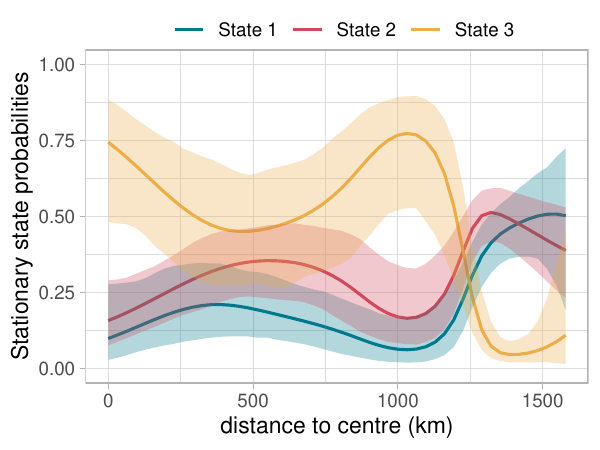} \caption{Transition probability $\gamma_{32}$ (left) and stationary state probabilities (right) as functions of distance to central location (in km), with 95\% pointwise confidence bands, for petrel analysis. The random intercept for individual \code{PET-A} is used.}\label{fig:d2c}
\end{figure}

\end{knitrout}

Figure \ref{fig:d2c} indicates that the fast ``travelling'' state was most probable for distances to central location smaller than 1000km, but this probability decreases sharply above 1000km. This suggests that petrels tend to travel at high speeds to reach good foraging grounds far from their central location, and slow down to forage (or search for food) when they reach those areas of interest. 

\subsubsection{Inter-individual heterogeneity}

We can visualise the inter-individual heterogeneity in the state process using the \code{HMM\$plot()} method, which shows the predicted values of the random effect with confidence intervals (Figure \ref{fig:ID}). This requires setting the other covariate(s) to a fixed value, and here we choose $d_t = 1500$ km.

\begin{knitrout}
\definecolor{shadecolor}{rgb}{0.969, 0.969, 0.969}\color{fgcolor}\begin{kframe}
\begin{alltt}
\hldef{hmm}\hlopt{$}\hlkwd{plot}\hldef{(}\hlkwc{what} \hldef{=} \hlsng{"delta"}\hldef{,} \hlkwc{var} \hldef{=} \hlsng{"ID"}\hldef{,} \hlkwc{covs} \hldef{=} \hlkwd{list}\hldef{(}\hlkwc{d2c} \hldef{=} \hlnum{1500}\hldef{))}
\end{alltt}
\end{kframe}\begin{figure}

{\centering \includegraphics[width=0.7\linewidth]{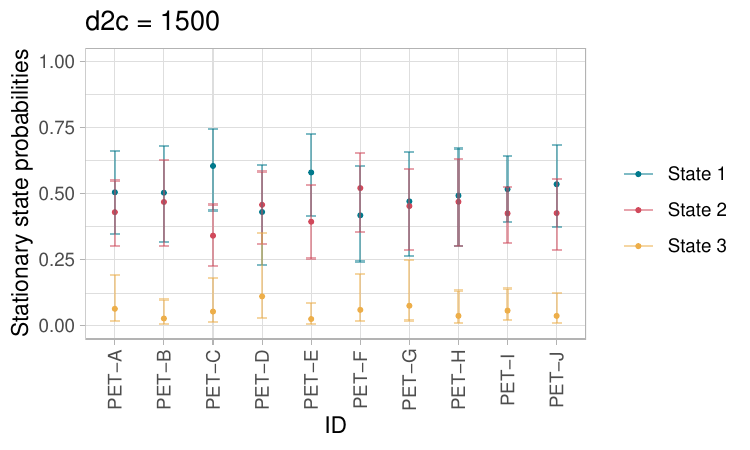} 

}

\caption[Predicted individual-specific random intercepts for petrel analysis, with 95\% confidence intervals]{Predicted individual-specific random intercepts for petrel analysis, with 95\% confidence intervals.}\label{fig:ID}
\end{figure}

\end{knitrout}

In many applications, these random intercepts are not of direct interest, and we would instead report their variance. The standard deviation of the random effects, returned by \code{HMM\$sd\_re()}, quantifies inter-individual variability.
\begin{knitrout}
\definecolor{shadecolor}{rgb}{0.969, 0.969, 0.969}\color{fgcolor}\begin{kframe}
\begin{alltt}
\hlcom{# "hid" gives the component for the hidden state model}
\hldef{hmm}\hlopt{$}\hlkwd{sd_re}\hldef{()}\hlopt{$}\hldef{hid}
\end{alltt}
\begin{verbatim}
                [,1]
S1>S2.s(ID)  0.29797
S1>S2.s(d2c) 0.01236
S2>S1.s(ID)  0.12672
S2>S1.s(d2c) 0.00172
S2>S3.s(ID)  0.45705
S2>S3.s(d2c) 0.12058
S3>S2.s(ID)  0.42037
S3>S2.s(d2c) 0.02568
\end{verbatim}
\end{kframe}
\end{knitrout}

The function returns two standard deviations for each transition probability: one for the random intercept (interpreted as the standard deviation of the normal distribution), and one for the spline (inversely related to its smoothness). The distributions of random intercepts were estimated to $\beta_{12}^{(k)} \sim N(0, 0.30^2)$, $\beta_{21}^{(k)} \sim N(0, 0.11^2)$, $\beta_{23}^{(k)} \sim N(0, 0.47^2)$, and $\beta_{32}^{(k)} \sim N(0, 0.41^2)$.

\section{Conclusion}

HMMs are widely applied in many fields, and we anticipate that \pkg{hmmTMB} will be of general interest among statisticians and practitioners. The flexible model formulation allowed by the package, including covariate effects and parameter constraints, makes it possible to implement many general Markov-switching models, placing it at the cutting edge of HMM research.

The package is based on \pkg{TMB} to take advantage of its great flexibility and computational tractability, but this comes with some pitfalls. In particular, the Laplace approximation is used to derive the marginal likelihood of the model, and this may perform poorly in cases where the likelihood is multimodal or asymmetric. The performance of the Laplace approximation for HMMs is an open question, and we hope that this can be investigated by comparing outputs from \code{HMM\$fit()} to outputs from \code{HMM\$fit\_stan()} (which samples random effects using \pkg{Stan}). It is also important to note that, regardless of the implementation method, complex model formulations (e.g., including multiple penalised splines) are susceptible to numerical instability and prohibitive computational cost. It is more challenging to estimate covariate effects in HMMs than in generalised linear models, for example, because parts of the model are latent and need to be inferred. Although \pkg{hmmTMB} makes it straightforward in principle to include many complex relationships in a model, we recommend parsimonious formulations.

The modular structure of \pkg{hmmTMB} makes it easy to extend, and we plan to include additional functionalities in the future. For example, we recognise that some important probability distributions are not currently included in the package, and we are open to request for additional distributions. Another important extension would be to implement continuous-time HMMs, which allow for observations at irregular time intervals, and where state transitions can occur at any time, making them popular in areas where sampling is irregular \citep{jackson2003}. It might also be possible to implement hierarchical HMMs within \pkg{hmmTMB} in the future, if such a model can be rewritten as an HMM over an expanded state space, similarly to the general dependence structures described in Section \ref{sec:general_dependence}. Hierarchical HMMs are based on two nested state processes, and they have recently been used to study the patterns of financial markets \citep{oelschlager2023} and the behaviour of animals \citep{leos2017multi} over multiple time scales.

\section*{Acknowledgements}

I am very grateful to Richard Glennie, who provided key ideas and code in the early development of this package. I would also like to thank Carlina Feldmann, Sina Mews, Brett McClintock, Vinky Wang, and Jan-Ole Koslik for helpful feedback about the package, and Natasha Klappstein for providing comments about an earlier version of this paper.

\bibliographystyle{apalike}
\bibliography{refs.bib}

\renewcommand\thefigure{S\arabic{figure}}
\setcounter{figure}{0}
\newpage
\begin{appendix}
  \section{List of distributions}
  \label{app:dists}
  
  In \pkg{hmmTMB}, the observation process is modelled with state-dependent (parametric) distributions,
  \begin{equation*}
    Z_t \mid \{ S_t = j \} \sim \mathcal{D}(\bm\omega_j)
  \end{equation*}
  where $j \in \{ 1, 2, \dots, K \}$. The choice of $\mathcal{D}$ depends on the type of quantity modelled by $Z_t$; e.g., a positive count might be modelled with Poisson distributions, a proportion between 0 and 1 might be modelled with beta distributions, and an unconstrained continuous variable might be modelled with normal distributions. 
  
  This is the list of distributions currently available in \pkg{hmmTMB}, with a list of parameters. It is relatively easy to add new distributions to the package, and users are invited to express interest in missing distributions, for future releases.
  
\begin{knitrout}
\definecolor{shadecolor}{rgb}{0.969, 0.969, 0.969}\color{fgcolor}\begin{kframe}
\begin{verbatim}
"beta": beta(shape1, shape2)
"binom": binomial(size, prob)
"cat": categorical(p2, p3, ...)
"dir": Dirichlet(alpha1, alpha2)
"exp": exponential(rate)
"foldednorm": folded normal(mean, sd)
"gamma": gamma(shape, scale)
"gamma2": gamma2(mean, sd)
"lnorm": log-normal(meanlog, sdlog)
"mvnorm": multivariate normal(mu1, mu2, ..., sd1, sd2, ..., corr12, ...)
"nbinom": negative binomial(size, prob)
"nbinom2": negative binomial(mean, shape)
"norm": normal(mean, sd)
"pois": Poisson(rate)
"t": Student's t(mean, scale)
"truncnorm": truncated normal(mean, sd, min, max)
"tweedie": Tweedie(mean, p, phi)
"vm": von Mises(mu, kappa)
"weibull": Weibull(shape, scale)
"wrpcauchy": wrapped Cauchy(mu, rho)
"zibinom": zero-inflated binomial(size, prob, z)
"zigamma": zero-inflated gamma(shape, scale, z)
"zigamma2": zero-inflated gamma2(mean, sd, z)
"zinbinom": zero-inflated negative binomial(size, prob, z)
"zipois": zero-inflated Poisson(rate, z)
"zoibeta": zoibeta(shape1, shape2, zeromass, onemass)
"ztnbinom": zero-truncated negative binomial(size, prob)
"ztpois": zero-truncated Poisson(rate)
\end{verbatim}
\end{kframe}
\end{knitrout}

The link functions used for the parameters are the following:
\begin{itemize}
\item log for $\omega > 0$
\item logit for $\omega \in [0, 1]$
\item logit of scaled parameter for $\omega \in [a, b]$ (e.g., mean of angular distribution $\in [-\pi, \pi]$)
\item identity for $\omega \in \mathbb{R}$
\end{itemize}

\section{Energy price analysis}
\label{app:energy}

In this appendix, we illustrate the implementation of Markov-switching regression in \pkg{hmmTMB}, with the inclusion of covariates in the observation parameters. We present the analysis of a data set of energy prices, included in the \proglang{R} package \pkg{MSwM} \citep{sanchez2021}. The data set includes energy prices in Spain between 2002 and 2008, as well as some potential explanatory variables (e.g., raw material prices, financial indices). Refer to \code{?MSwM::energy} for more detail about the data set.

\begin{knitrout}
\definecolor{shadecolor}{rgb}{0.969, 0.969, 0.969}\color{fgcolor}\begin{kframe}
\begin{alltt}
\hlkwd{library}\hldef{(hmmTMB)}
\hlkwd{data}\hldef{(energy,} \hlkwc{package} \hldef{=} \hlsng{"MSwM"}\hldef{)}
\end{alltt}
\end{kframe}
\end{knitrout}

\subsection{Model formulation}

We consider a 2-state model, where the energy price $Z_t$ is modelled with a normal distribution in each state,
\begin{equation*}
  Z_t \mid S_t = j \sim N(\mu_j, \sigma_j)
\end{equation*}

We further make the assumption that, within each state, the parameters of the normal distribution depend on the exchange rate between euro and US dollar, $r_t$. The mean is modelled with a cubic spline, and the standard deviation is modelled with a cubic polynomial.
\begin{align*}
  \mu_j^{(t)} & = \alpha_0 + f_j(r_t) \\
  \log(\sigma_j^{(t)}) & = \alpha_0 + \alpha_1 r_t + \alpha_2 r_t^2 + \alpha_3 r_t^3
\end{align*}
This is an example of Markov-switching generalised additive model.

\subsection{Model fitting}

The hidden state model does not include covariates in this analysis so, to specify the \code{MarkovChain} object, we only need to pass the data frame (needed to create design matrices, even when there are no covariates) and the number of states.
\begin{knitrout}
\definecolor{shadecolor}{rgb}{0.969, 0.969, 0.969}\color{fgcolor}\begin{kframe}
\begin{alltt}
\hlcom{# Create hidden state model}
\hldef{hid} \hlkwb{<-} \hldef{MarkovChain}\hlopt{$}\hlkwd{new}\hldef{(}\hlkwc{data} \hldef{= energy,} \hlkwc{n_states} \hldef{=} \hlnum{2}\hldef{)}
\end{alltt}
\end{kframe}
\end{knitrout}

Defining the observation model involves a little more work, as it requires a list of observation distributions, a list of initial parameters, and a list of model formulas. We choose initial parameters based on inspection of a histogram of the \code{Price} variable. We use syntax from the \proglang{R} package \pkg{mgcv} to define non-parametric formula terms.

\begin{knitrout}
\definecolor{shadecolor}{rgb}{0.969, 0.969, 0.969}\color{fgcolor}\begin{kframe}
\begin{alltt}
\hlcom{# List of observation distributions}
\hldef{dists} \hlkwb{<-} \hlkwd{list}\hldef{(}\hlkwc{Price} \hldef{=} \hlsng{"norm"}\hldef{)}
\hlcom{# List of initial parameters}
\hldef{par0} \hlkwb{<-} \hlkwd{list}\hldef{(}\hlkwc{Price} \hldef{=} \hlkwd{list}\hldef{(}\hlkwc{mean} \hldef{=} \hlkwd{c}\hldef{(}\hlnum{3}\hldef{,} \hlnum{6}\hldef{),} \hlkwc{sd} \hldef{=} \hlkwd{c}\hldef{(}\hlnum{1}\hldef{,} \hlnum{1}\hldef{)))}
\hlcom{# List of formulas}
\hldef{f} \hlkwb{<-} \hlkwd{list}\hldef{(}\hlkwc{Price} \hldef{=} \hlkwd{list}\hldef{(}\hlkwc{mean} \hldef{=} \hlopt{~} \hlkwd{s}\hldef{(EurDol,} \hlkwc{k} \hldef{=} \hlnum{10}\hldef{,} \hlkwc{bs} \hldef{=} \hlsng{"cs"}\hldef{),}
                       \hlkwc{sd} \hldef{=} \hlopt{~} \hlkwd{poly}\hldef{(EurDol,} \hlnum{3}\hldef{)))}

\hlcom{# Create observation model}
\hldef{obs} \hlkwb{<-} \hldef{Observation}\hlopt{$}\hlkwd{new}\hldef{(}\hlkwc{data} \hldef{= energy,}
                       \hlkwc{n_states} \hldef{=} \hlnum{2}\hldef{,}
                       \hlkwc{dists} \hldef{= dists,}
                       \hlkwc{par} \hldef{= par0,}
                       \hlkwc{formulas} \hldef{= f)}
\end{alltt}
\end{kframe}
\end{knitrout}

We combine the \code{MarkovChain} and \code{Observation} objects to create the model, and we fit it. This takes about 50 sec on a laptop with an Intel i7-1065G7 CPU @1.30GHz with 16Gb RAM.
\begin{knitrout}
\definecolor{shadecolor}{rgb}{0.969, 0.969, 0.969}\color{fgcolor}\begin{kframe}
\begin{alltt}
\hldef{hmm} \hlkwb{<-} \hldef{HMM}\hlopt{$}\hlkwd{new}\hldef{(}\hlkwc{hid} \hldef{= hid,} \hlkwc{obs} \hldef{= obs)}
\hldef{hmm}\hlopt{$}\hlkwd{fit}\hldef{(}\hlkwc{silent} \hldef{=} \hlnum{TRUE}\hldef{)}
\end{alltt}
\end{kframe}
\end{knitrout}

\subsection{Visualise the results}

The main object of interest in this analysis is the relationship between the observation parameters and the covariate. We can visualise this with the method \code{HMM\_plot()}, with the arguments \code{what = "obspar"} and \code{var = "EurDol"} to indicate that we want to visualise the observation parameters as functions of the euro-dollar exchange rate covariate. We can also use the \code{i} argument to output only the plot of the mean or the plot of the standard deviation. Here, we want to overlay the relationship between mean price and euro-dollar exchange rate with the data. As \code{HMM\_plot()} returns a ggplot object, we can add the points with the usual ggplot syntax. We colour them by the most likely state sequence from the Viterbi algorithm, to visualise how observations were classified by the model. We use the same method to visualise the standard deviation of price.
\begin{knitrout}
\definecolor{shadecolor}{rgb}{0.969, 0.969, 0.969}\color{fgcolor}\begin{kframe}
\begin{alltt}
\hlcom{# Get most likely state sequence for plotting}
\hldef{energy}\hlopt{$}\hldef{viterbi} \hlkwb{<-} \hlkwd{factor}\hldef{(}\hlkwd{paste0}\hldef{(}\hlsng{"State "}\hldef{, hmm}\hlopt{$}\hlkwd{viterbi}\hldef{()))}

\hlcom{# Plot mean price in each state, with data points}
\hldef{hmm}\hlopt{$}\hlkwd{plot}\hldef{(}\hlkwc{what} \hldef{=} \hlsng{"obspar"}\hldef{,} \hlkwc{var} \hldef{=} \hlsng{"EurDol"}\hldef{,} \hlkwc{i} \hldef{=} \hlsng{"Price.mean"}\hldef{)} \hlopt{+}
    \hlkwd{geom_point}\hldef{(}\hlkwd{aes}\hldef{(}\hlkwc{x} \hldef{= EurDol,} \hlkwc{y} \hldef{= Price,} \hlkwc{fill} \hldef{= viterbi,} \hlkwc{col} \hldef{= viterbi),}
               \hlkwc{data} \hldef{= energy,} \hlkwc{alpha} \hldef{=} \hlnum{0.3}\hldef{)} \hlopt{+}
    \hlkwd{theme}\hldef{(}\hlkwc{legend.position} \hldef{=} \hlsng{"none"}\hldef{)}

\hlcom{# Plot price standard deviation in each state}
\hldef{hmm}\hlopt{$}\hlkwd{plot}\hldef{(}\hlkwc{what} \hldef{=} \hlsng{"obspar"}\hldef{,} \hlkwc{var} \hldef{=} \hlsng{"EurDol"}\hldef{,} \hlkwc{i} \hldef{=} \hlsng{"Price.sd"}\hldef{)} \hlopt{+}
    \hlkwd{theme}\hldef{(}\hlkwc{legend.position} \hldef{=} \hlkwd{c}\hldef{(}\hlnum{0.3}\hldef{,} \hlnum{0.7}\hldef{))}
\end{alltt}
\end{kframe}
\includegraphics[width=0.49\linewidth]{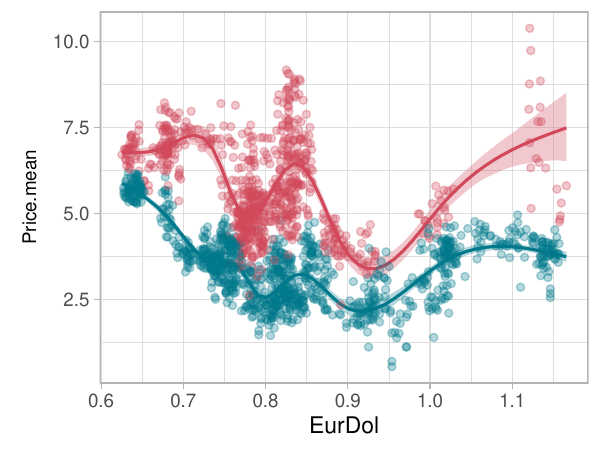} 
\includegraphics[width=0.49\linewidth]{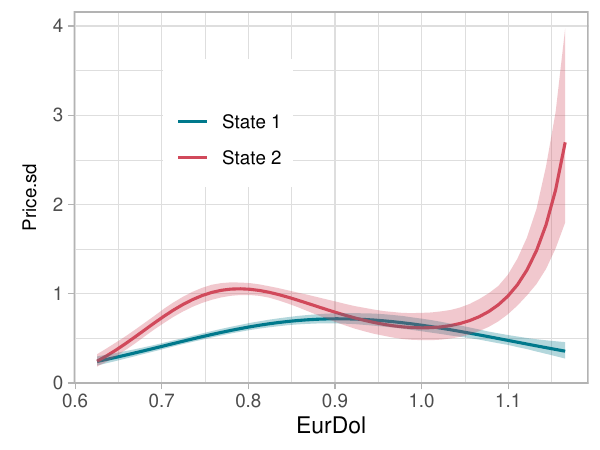} 
\end{knitrout}

Overall, state 1 captured lower energy prices, and state 2 higher energy prices, although the mean price in each state varied greatly with the euro-dollar exchange rate. In both states, the relationship between mean price and exchange rate was highly non-linear. There also seemed to be a clear effect of the exchange rate on the standard deviation of price, i.e., its variability.

\subsection{Predict state-dependent distributions}

In many studies, it is interesting to plot the state-dependent density functions, possibly on top of a histogram of the data. This can for example be helpful to interpret the states, to assess how much they overlap, or to determine whether the estimated distributions capture the distribution of the data. In this example, the density function in each state depends on the exchange rate variable, and so we can only create the plot for a chosen value of this covariate. Here, we do this for a few different values of the exchange rate, to visualise how the distributions change.

We first create a data frame of the euro-dollar exchange rate values for which the distributions should be computed; here, we choose six values that roughly span the range of the variable. We pass this data frame to \code{HMM\$predict()} to get an array of the state-dependent observation parameters, calculated for those covariate values.
\begin{knitrout}
\definecolor{shadecolor}{rgb}{0.969, 0.969, 0.969}\color{fgcolor}\begin{kframe}
\begin{alltt}
\hlcom{# Get state-dependent parameters for a few values of EurDol}
\hldef{EurDol} \hlkwb{<-} \hlkwd{seq}\hldef{(}\hlnum{0.65}\hldef{,} \hlnum{1.15}\hldef{,} \hlkwc{by} \hldef{=} \hlnum{0.1}\hldef{)}
\hldef{newdata} \hlkwb{<-} \hlkwd{data.frame}\hldef{(}\hlkwc{EurDol} \hldef{= EurDol)}
\hldef{par} \hlkwb{<-} \hldef{hmm}\hlopt{$}\hlkwd{predict}\hldef{(}\hlkwc{what} \hldef{=} \hlsng{"obspar"}\hldef{,} \hlkwc{newdata} \hldef{= newdata)}
\end{alltt}
\end{kframe}
\end{knitrout}

In preparation to create the plots, we derive the proportion of time spent in each state (as determined by the Viterbi state sequence), which will be used as a weight for each state-dependent density. We also create a grid of values of energy prices, which will be shown on the x axis of the plots.
\begin{knitrout}
\definecolor{shadecolor}{rgb}{0.969, 0.969, 0.969}\color{fgcolor}\begin{kframe}
\begin{alltt}
\hlcom{# Weights for state-dependent distributions}
\hldef{w} \hlkwb{<-} \hlkwd{table}\hldef{(hmm}\hlopt{$}\hlkwd{viterbi}\hldef{())}\hlopt{/}\hlkwd{nrow}\hldef{(energy)}

\hlcom{# Grid of energy prices (x axis)}
\hldef{grid} \hlkwb{<-} \hlkwd{seq}\hldef{(}\hlkwd{min}\hldef{(energy}\hlopt{$}\hldef{Price),} \hlkwd{max}\hldef{(energy}\hlopt{$}\hldef{Price),} \hlkwc{length} \hldef{=} \hlnum{100}\hldef{)}
\end{alltt}
\end{kframe}
\end{knitrout}

We compute the state-dependent probability densities using \code{dnorm()}, by plugging in the predict parameter values, and we format them into a data frame for plotting.
\begin{knitrout}
\definecolor{shadecolor}{rgb}{0.969, 0.969, 0.969}\color{fgcolor}\begin{kframe}
\begin{alltt}
\hlcom{# For each value of EurDol, compute state-dependent distributions}
\hldef{pdf_ls} \hlkwb{<-} \hlkwd{lapply}\hldef{(}\hlnum{1}\hlopt{:}\hlkwd{dim}\hldef{(par)[}\hlnum{3}\hldef{],} \hlkwa{function}\hldef{(}\hlkwc{i}\hldef{) \{}
  \hldef{p} \hlkwb{<-} \hldef{par[,,i]}
  \hldef{pdf1} \hlkwb{<-} \hldef{w[}\hlnum{1}\hldef{]} \hlopt{*} \hlkwd{dnorm}\hldef{(grid,} \hlkwc{mean} \hldef{= p[}\hlsng{"Price.mean"}\hldef{,} \hlsng{"state 1"}\hldef{],}
                       \hlkwc{sd} \hldef{= p[}\hlsng{"Price.sd"}\hldef{,} \hlsng{"state 1"}\hldef{])}
  \hldef{pdf2} \hlkwb{<-} \hldef{w[}\hlnum{2}\hldef{]} \hlopt{*} \hlkwd{dnorm}\hldef{(grid,} \hlkwc{mean} \hldef{= p[}\hlsng{"Price.mean"}\hldef{,} \hlsng{"state 2"}\hldef{],}
                       \hlkwc{sd} \hldef{= p[}\hlsng{"Price.sd"}\hldef{,} \hlsng{"state 2"}\hldef{])}
  \hldef{res} \hlkwb{<-} \hlkwd{data.frame}\hldef{(}\hlkwc{price} \hldef{= grid,} \hlkwc{pdf} \hldef{=} \hlkwd{c}\hldef{(pdf1, pdf2),}
                    \hlkwc{state} \hldef{=} \hlkwd{factor}\hldef{(}\hlkwd{rep}\hldef{(}\hlnum{1}\hlopt{:}\hlnum{2}\hldef{,} \hlkwc{each} \hldef{=} \hlkwd{length}\hldef{(grid))),}
                    \hlkwc{eurdol} \hldef{=} \hlkwd{paste0}\hldef{(}\hlsng{"EurDol = "}\hldef{, EurDol[i]))}
  \hlkwd{return}\hldef{(res)}
\hldef{\})}
\hlcom{# Turn into dataframe for ggplot}
\hldef{pdf_df} \hlkwb{<-} \hlkwd{do.call}\hldef{(rbind, pdf_ls)}
\end{alltt}
\end{kframe}
\end{knitrout}

Finally, we plot histograms of the observed energy prices, and add the state-dependent normal density lines, for the different values of exchange rate. The distributions vary widely for different values of the covariate.
\begin{knitrout}
\definecolor{shadecolor}{rgb}{0.969, 0.969, 0.969}\color{fgcolor}\begin{kframe}
\begin{alltt}
\hlkwd{ggplot}\hldef{(pdf_df,} \hlkwd{aes}\hldef{(price, pdf))} \hlopt{+}
  \hlkwd{facet_wrap}\hldef{(}\hlsng{"eurdol"}\hldef{)} \hlopt{+}
  \hlkwd{geom_histogram}\hldef{(}\hlkwd{aes}\hldef{(}\hlkwc{x} \hldef{= Price,} \hlkwc{y}\hldef{=..density..),} \hlkwc{bins} \hldef{=} \hlnum{20}\hldef{,}
                 \hlkwc{col} \hldef{=} \hlsng{"white"}\hldef{,} \hlkwc{bg} \hldef{=} \hlsng{"lightgrey"}\hldef{,} \hlkwc{data} \hldef{= energy)} \hlopt{+}
  \hlkwd{geom_line}\hldef{(}\hlkwd{aes}\hldef{(}\hlkwc{col} \hldef{= state))} \hlopt{+}
  \hlkwd{theme_bw}\hldef{()} \hlopt{+}
  \hlkwd{labs}\hldef{(}\hlkwc{x} \hldef{=} \hlsng{"Price"}\hldef{,} \hlkwc{y} \hldef{=} \hlkwa{NULL}\hldef{)} \hlopt{+}
  \hlkwd{scale_color_manual}\hldef{(}\hlkwc{values} \hldef{= hmmTMB}\hlopt{:::}\hldef{hmmTMB_cols)}
\end{alltt}
\end{kframe}

{\centering \includegraphics[width=1\linewidth]{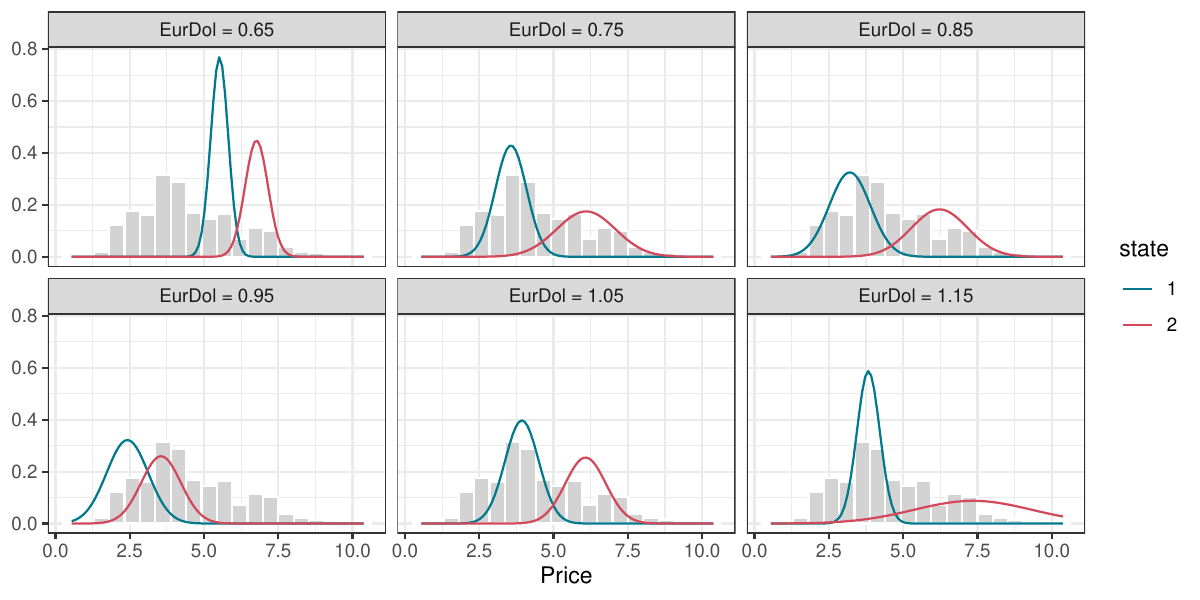} 

}

\end{knitrout}

\section{Bayesian analysis of human activity data}
\label{app:activ}

We analyse a subset of the data described by \cite{huang2018}, accessible on Github at \url{github.com/huang1010/hmms}. In that study, accelerometers were attached to human subjects to measure their activity through time. Raw acceleration was processed to obtain continuous variable summarising physical activity at a 5-min resolution. We consider data from two subjects, named ``S9'' and ``S20'', comprising a total of 2317 observations over around 8 days.

The aim of the study was to investigate circadian rythms, i.e., daily patterns in activity. Similarly to \cite{huang2018}, we included time of day as a covariate on the transition probabilities but, unlike them, we used a non-parametric approach. Specifically, we modelled the relationship between transition probabilities and time of day using a cyclic spline (\code{"cc"} in \pkg{mgcv}), with period set to 24 hours.

We follow a Bayesian approach for this analysis, i.e., we perform posterior inference using \code{HMM\$fit\_stan()}. 

\subsection{Prepare data}

We load the data from Github, save them in the format expected by \pkg{hmmTMB}, and add a column for time of day (numeric between 0 and 24). Plotting the time series of activity measurements suggests that there are some clearly distinct phases of low and high activity. We will use a 2-state HMM to try to identify those phases.

\begin{knitrout}
\definecolor{shadecolor}{rgb}{0.969, 0.969, 0.969}\color{fgcolor}\begin{kframe}
\begin{alltt}
\hlcom{# Load data from Github URL}
\hldef{path} \hlkwb{<-} \hlsng{"https://raw.githubusercontent.com/huang1010/HMMS/master/"}
\hldef{data1} \hlkwb{<-} \hlkwd{read.csv}\hldef{(}\hlkwd{url}\hldef{(}\hlkwd{paste0}\hldef{(path,} \hlsng{"S9.csv"}\hldef{)))[,}\hlopt{-}\hlnum{3}\hldef{]}
\hldef{data2} \hlkwb{<-} \hlkwd{read.csv}\hldef{(}\hlkwd{url}\hldef{(}\hlkwd{paste0}\hldef{(path,} \hlsng{"S20.csv"}\hldef{)))[,}\hlopt{-}\hlnum{3}\hldef{]}

\hlcom{# Format data }
\hldef{data} \hlkwb{<-} \hlkwd{rbind}\hldef{(}\hlkwd{cbind}\hldef{(}\hlkwc{ID} \hldef{=} \hlsng{"S9"}\hldef{, data1),}
              \hlkwd{cbind}\hldef{(}\hlkwc{ID} \hldef{=} \hlsng{"S20"}\hldef{, data2))}
\hldef{data}\hlopt{$}\hldef{time} \hlkwb{<-} \hlkwd{as.POSIXct}\hldef{(data}\hlopt{$}\hldef{time)}
\hlcom{# Get time of day}
\hldef{data}\hlopt{$}\hldef{tod} \hlkwb{<-} \hlkwd{as.numeric}\hldef{(}\hlkwd{format}\hldef{(data}\hlopt{$}\hldef{time,} \hlsng{"%H"}\hldef{))} \hlopt{+}
    \hlkwd{as.numeric}\hldef{(}\hlkwd{format}\hldef{(data}\hlopt{$}\hldef{time,} \hlsng{"%M"}\hldef{))}\hlopt{/}\hlnum{60}

\hlcom{# Plot activity against time for each individual}
\hlkwd{theme_set}\hldef{(}\hlkwd{theme_bw}\hldef{())}
\hlkwd{ggplot}\hldef{(data,} \hlkwd{aes}\hldef{(time, activity))} \hlopt{+}
    \hlkwd{facet_wrap}\hldef{(}\hlsng{"ID"}\hldef{,} \hlkwc{scales} \hldef{=} \hlsng{"free_x"}\hldef{)} \hlopt{+}
    \hlkwd{geom_line}\hldef{()}
\end{alltt}
\end{kframe}

{\centering \includegraphics[width=1\linewidth]{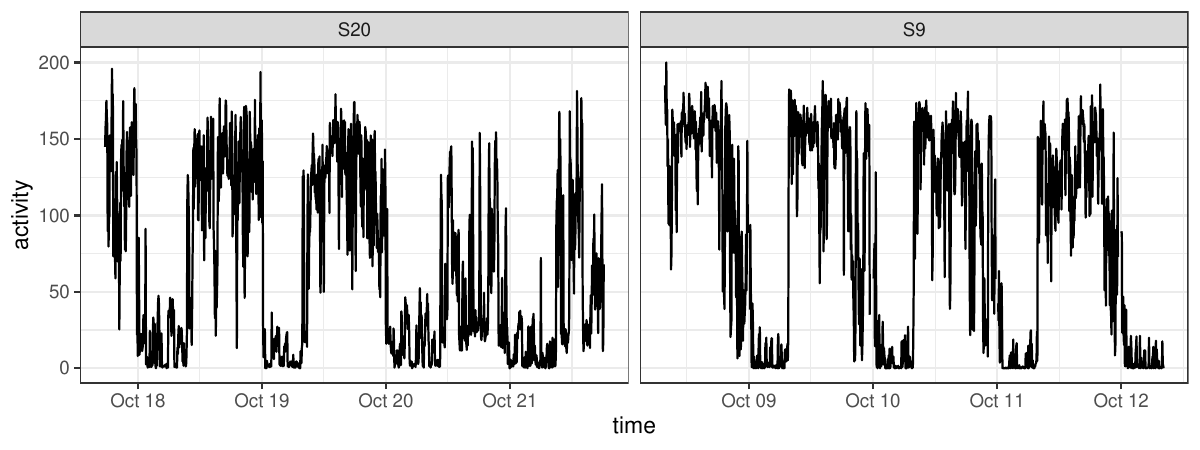} 

}

\end{knitrout}

\subsection{Model specification}

We use a cyclic spline for time of day to capture its effect on the transition probabilities. We use \code{initial\_state = "stationary"} to indicate that the initial distribution of the Markov chain should be fixed to the stationary distribution of the first transition probability matrix, rather than estimated. This is to avoid mixing problems arising from poor identifiability of the initial distribution.

For the observation model, we use the \code{"zigamma2"} distribution, i.e., the zero-inflated gamma distribution specified in terms of mean and standard deviation (rather than shape and scale). We need the zero-inflated distribution because a small number of activity values are exactly zero, and those cannot be modelled with a gamma distribution. As a result, an additional parameter \code{z} is required, to measure the proportion of values that are exactly zero in each state. We choose initial values based on the expectation that state 1 will capture active behaviour, and state 2 will capture less active periods. 

In this example, we fix the zero inflation parameter for state 1 to zero, i.e., we assume that state 2 will capture all observations that are exactly zero. We do this because initial experiments revealed that the zero inflation parameter for state 1 was estimated to be virtually zero, and mixing was poor.

\begin{knitrout}
\definecolor{shadecolor}{rgb}{0.969, 0.969, 0.969}\color{fgcolor}\begin{kframe}
\begin{alltt}
\hlcom{# Define Markov chain model}
\hldef{hid} \hlkwb{<-} \hldef{MarkovChain}\hlopt{$}\hlkwd{new}\hldef{(}\hlkwc{data} \hldef{= data,} \hlkwc{n_states} \hldef{=} \hlnum{2}\hldef{,} \hlkwc{initial_state} \hldef{=} \hlsng{"stationary"}\hldef{,}
                       \hlkwc{formula} \hldef{=} \hlopt{~}\hlkwd{s}\hldef{(tod,} \hlkwc{k} \hldef{=} \hlnum{5}\hldef{,} \hlkwc{bs} \hldef{=} \hlsng{"cc"}\hldef{))}

\hlcom{# Define observation model}
\hldef{dists} \hlkwb{<-} \hlkwd{list}\hldef{(}\hlkwc{activity} \hldef{=} \hlsng{"zigamma2"}\hldef{)}
\hldef{par0} \hlkwb{<-} \hlkwd{list}\hldef{(}\hlkwc{activity} \hldef{=} \hlkwd{list}\hldef{(}\hlkwc{mean} \hldef{=} \hlkwd{c}\hldef{(}\hlnum{20}\hldef{,} \hlnum{150}\hldef{),}
                             \hlkwc{sd} \hldef{=} \hlkwd{c}\hldef{(}\hlnum{20}\hldef{,} \hlnum{40}\hldef{),}
                             \hlkwc{z} \hldef{=} \hlkwd{c}\hldef{(}\hlnum{0.1}\hldef{,} \hlnum{0}\hldef{)))}
\hldef{obs} \hlkwb{<-} \hldef{Observation}\hlopt{$}\hlkwd{new}\hldef{(}\hlkwc{data} \hldef{= data,} \hlkwc{dists} \hldef{= dists,}
                       \hlkwc{n_states} \hldef{=} \hlnum{2}\hldef{,} \hlkwc{par} \hldef{= par0)}

\hlcom{# Fixed parameter (zero mass in state 2)}
\hldef{fixpar} \hlkwb{<-} \hlkwd{list}\hldef{(}\hlkwc{obs} \hldef{=} \hlkwd{c}\hldef{(}\hlsng{"activity.z.state2.(Intercept)"} \hldef{=} \hlnum{NA}\hldef{))}

\hlcom{# Define HMM}
\hldef{hmm} \hlkwb{<-} \hldef{HMM}\hlopt{$}\hlkwd{new}\hldef{(}\hlkwc{obs} \hldef{= obs,} \hlkwc{hid} \hldef{= hid,} \hlkwc{fixpar} \hldef{= fixpar)}
\end{alltt}
\end{kframe}
\end{knitrout}

We also specify prior distributions for some of the parameters using \code{HMM\$set\_priors()}. We pass a list with elements \code{coeff\_fe\_obs} and \code{coeff\_fe\_hid}, for the fixed effect parameters of the observation model and the hidden state model, respectively. For the observation model, we provide a matrix with six rows, corresponding to intercepts for the mean in state 1, mean in state 2, standard deviation in state 1, standard deviation in state 2, zero mass in state 1, and zero mass in state 2. The matrix has one column for the mean of the normal prior distribution, and one column for its standard deviation. We use diffuse priors centred on our initial guesses (with standard deviation 10), except for the zero mass in state 2, which is fixed to zero and not estimated. For the hidden state process, we specify a matrix with one row each for the intercepts of $\Pr(S_{t+1} = 2 \mid S_t = i)$ and $\Pr(S_{t+1} = 1 \mid S_t = 2)$, and we choose these to represent our expectation that the off-diagonal elements of the transition probability matrix will be small. (This is common because the state process tends to display persistence.) We could also specify priors for the smoothness parameters of the penalised splines, but we leave them as \code{NA} here (i.e., improper flat priors are used).

\begin{knitrout}
\definecolor{shadecolor}{rgb}{0.969, 0.969, 0.969}\color{fgcolor}\begin{kframe}
\begin{alltt}
\hldef{prior_obs} \hlkwb{<-} \hlkwd{matrix}\hldef{(}\hlkwd{c}\hldef{(}\hlkwd{log}\hldef{(}\hlnum{20}\hldef{),} \hlnum{10}\hldef{,}
                     \hlkwd{log}\hldef{(}\hlnum{150}\hldef{),} \hlnum{10}\hldef{,}
                     \hlkwd{log}\hldef{(}\hlnum{20}\hldef{),} \hlnum{10}\hldef{,}
                     \hlkwd{log}\hldef{(}\hlnum{40}\hldef{),} \hlnum{10}\hldef{,}
                     \hlkwd{qlogis}\hldef{(}\hlnum{0.1}\hldef{),} \hlnum{10}\hldef{,}
                     \hlnum{NA}\hldef{,} \hlnum{NA}\hldef{),}
                   \hlkwc{ncol} \hldef{=} \hlnum{2}\hldef{,} \hlkwc{byrow} \hldef{=} \hlnum{TRUE}\hldef{)}
\hldef{prior_hid} \hlkwb{<-} \hlkwd{matrix}\hldef{(}\hlkwd{c}\hldef{(}\hlkwd{qlogis}\hldef{(}\hlnum{0.05}\hldef{),} \hlnum{10}\hldef{,}
                      \hlkwd{qlogis}\hldef{(}\hlnum{0.05}\hldef{),} \hlnum{10}\hldef{),}
                    \hlkwc{ncol} \hldef{=} \hlnum{2}\hldef{,} \hlkwc{byrow} \hldef{=} \hlnum{TRUE}\hldef{)}

\hldef{hmm}\hlopt{$}\hlkwd{set_priors}\hldef{(}\hlkwd{list}\hldef{(}\hlkwc{coeff_fe_obs} \hldef{= prior_obs,}
                    \hlkwc{coeff_fe_hid} \hldef{= prior_hid))}

\hldef{hmm}\hlopt{$}\hlkwd{priors}\hldef{()}
\end{alltt}
\begin{verbatim}
$coeff_fe_obs
                                  mean sd
activity.mean.state1.(Intercept)  3.00 10
activity.mean.state2.(Intercept)  5.01 10
activity.sd.state1.(Intercept)    3.00 10
activity.sd.state2.(Intercept)    3.69 10
activity.z.state1.(Intercept)    -2.20 10
activity.z.state2.(Intercept)       NA NA

$coeff_fe_hid
                   mean sd
S1>S2.(Intercept) -2.94 10
S2>S1.(Intercept) -2.94 10

$log_lambda_obs
     mean sd

$log_lambda_hid
             mean sd
S1>S2.s(tod)   NA NA
S2>S1.s(tod)   NA NA
\end{verbatim}
\end{kframe}
\end{knitrout}

\subsection{Model fitting using Stan}

We run Hamiltonian Monte Carlo using \pkg{Stan}, with the function \code{HMM\$fit\_stan()}. We need to specify the number of chains and the number of iterations of each chain. Here, we run 1000 iterations for one chain; in practice, this choice should be made based on criteria such as effective sample sizes. This takes around 5 min on a laptop with an Intel i7-1065G7 CPU @1.30GHz with 16Gb RAM.

\begin{knitrout}
\definecolor{shadecolor}{rgb}{0.969, 0.969, 0.969}\color{fgcolor}\begin{kframe}
\begin{alltt}
\hldef{n_chain} \hlkwb{<-} \hlnum{1}
\hldef{n_iter} \hlkwb{<-} \hlnum{1000}
\hldef{hmm}\hlopt{$}\hlkwd{fit_stan}\hldef{(}\hlkwc{chains} \hldef{= n_chain,} \hlkwc{iter} \hldef{= n_iter)}
\end{alltt}
\end{kframe}
\end{knitrout}

After model fitting, \code{HMM\$iters()} returns a data frame of posterior samples, which can then be used to create trace plots or density plots, for example. By default, it outputs posterior samples for the HMM parameters (transition probabilities and observation parameters), but posterior samples for the linear predictor parameters (fixed effects, random effects, smoothness parameters) can be obtained using the argument \code{type = "raw"}. These data frames can also directly be passed to functions from the \pkg{bayesplot} package for visualisation, e.g., \code{bayesplot::mcmc\_pairs()} or \code{bayesplot::mcmc\_areas()}.

\begin{knitrout}
\definecolor{shadecolor}{rgb}{0.969, 0.969, 0.969}\color{fgcolor}\begin{kframe}
\begin{alltt}
\hldef{par_df} \hlkwb{<-} \hlkwd{as.data.frame.table}\hldef{(hmm}\hlopt{$}\hlkwd{iters}\hldef{()[,}\hlkwd{c}\hldef{(}\hlnum{1}\hlopt{:}\hlnum{4}\hldef{)])}
\hlkwd{colnames}\hldef{(par_df)} \hlkwb{<-} \hlkwd{c}\hldef{(}\hlsng{"iter"}\hldef{,} \hlsng{"par"}\hldef{,} \hlsng{"value"}\hldef{)}
\hlkwd{ggplot}\hldef{(par_df,} \hlkwd{aes}\hldef{(}\hlkwc{x} \hldef{= value))} \hlopt{+}
  \hlkwd{geom_histogram}\hldef{(}\hlkwc{bins} \hldef{=} \hlnum{30}\hldef{,} \hlkwc{col} \hldef{=} \hlsng{"grey"}\hldef{,} \hlkwc{fill} \hldef{=} \hlsng{"lightgrey"}\hldef{)} \hlopt{+}
  \hlkwd{facet_wrap}\hldef{(}\hlsng{"par"}\hldef{,} \hlkwc{scales} \hldef{=} \hlsng{"free"}\hldef{,} \hlkwc{ncol} \hldef{=} \hlnum{2}\hldef{)} \hlopt{+}
  \hlkwd{labs}\hldef{(}\hlkwc{x} \hldef{=} \hlkwa{NULL}\hldef{)}
\end{alltt}
\end{kframe}

{\centering \includegraphics[width=0.8\linewidth]{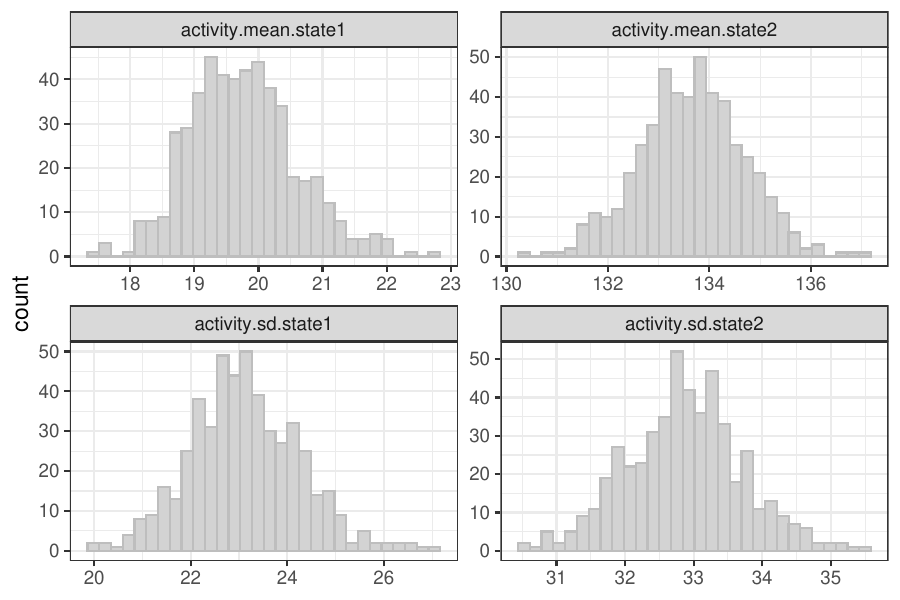} 

}

\end{knitrout}

We can also access the \code{stanfit} object using \code{HMM\$out\_stan()}, which can be printed, or passed to functions such as \code{rstan::stan\_trace()} or \code{rstan::stan\_dens()}. Printing the output gives posterior quantiles, as well as information about effective sample size and convergence. Here, \code{Rhat} is 1 for all parameters, indicating convergence of the sampler to the posterior distribution (see \pkg{rstan} and \pkg{Stan} documentation).

\begin{knitrout}
\definecolor{shadecolor}{rgb}{0.969, 0.969, 0.969}\color{fgcolor}\begin{kframe}
\begin{alltt}
\hldef{hmm}\hlopt{$}\hlkwd{out_stan}\hldef{()}
\end{alltt}
\begin{verbatim}
Inference for Stan model: hmmTMB.
1 chains, each with iter=1000; warmup=500; thin=1; 
post-warmup draws per chain=500, total post-warmup draws=500.

                       mean se_mean   sd      2.5%       25%       50%
coeff_fe_obs[1]        2.98    0.00 0.04      2.90      2.95      2.98
coeff_fe_obs[2]        4.90    0.00 0.01      4.88      4.89      4.90
coeff_fe_obs[3]        3.14    0.00 0.05      3.04      3.11      3.14
coeff_fe_obs[4]        3.49    0.00 0.03      3.44      3.48      3.49
coeff_fe_obs[5]       -2.30    0.00 0.10     -2.49     -2.37     -2.29
coeff_fe_hid[1]       -3.20    0.01 0.20     -3.58     -3.33     -3.20
coeff_fe_hid[2]       -3.40    0.01 0.26     -3.97     -3.55     -3.39
log_lambda_hid[1]      0.55    0.05 1.01     -1.81     -0.05      0.62
log_lambda_hid[2]      0.85    0.08 1.19     -1.67      0.12      0.95
coeff_re_hid[1]       -1.34    0.02 0.46     -2.29     -1.64     -1.32
coeff_re_hid[2]        0.69    0.02 0.44     -0.15      0.38      0.69
coeff_re_hid[3]        1.28    0.01 0.34      0.66      1.06      1.26
coeff_re_hid[4]       -0.44    0.04 0.58     -1.85     -0.75     -0.36
coeff_re_hid[5]       -1.38    0.02 0.39     -2.14     -1.64     -1.37
coeff_re_hid[6]       -0.57    0.02 0.37     -1.33     -0.79     -0.56
lp__              -10378.91    0.22 3.00 -10385.72 -10380.46 -10378.51
                        75%     97.5% n_eff Rhat
coeff_fe_obs[1]        3.01      3.07   422    1
coeff_fe_obs[2]        4.90      4.91  1013    1
coeff_fe_obs[3]        3.17      3.24   436    1
coeff_fe_obs[4]        3.51      3.54   709    1
coeff_fe_obs[5]       -2.23     -2.09   881    1
coeff_fe_hid[1]       -3.06     -2.82   690    1
coeff_fe_hid[2]       -3.21     -2.94   297    1
log_lambda_hid[1]      1.21      2.50   475    1
log_lambda_hid[2]      1.70      2.94   248    1
coeff_re_hid[1]       -1.03     -0.48   636    1
coeff_re_hid[2]        0.99      1.55   697    1
coeff_re_hid[3]        1.50      1.98   753    1
coeff_re_hid[4]       -0.05      0.47   262    1
coeff_re_hid[5]       -1.11     -0.66   384    1
coeff_re_hid[6]       -0.33      0.14   398    1
lp__              -10376.77 -10374.47   187    1

Samples were drawn using NUTS(diag_e) at Wed May 21 11:51:24 2025.
For each parameter, n_eff is a crude measure of effective sample size,
and Rhat is the potential scale reduction factor on split chains (at 
convergence, Rhat=1).
\end{verbatim}
\end{kframe}
\end{knitrout}

\subsection{Observation distributions}

It is often useful to visualise the state-dependent observation distributions, possibly combined with a histogram of the data, to help with state interpretation and assess goodness-of-fit. We can generate posterior samples of the distributions, and plot those to also visualise the associated uncertainty. The plot shows that the two states were clearly distinct: one of them captures high activity, and the other low activity. There seems to be some lack of fit for high activity values (i.e., the density lines don't match the distribution of observed data), and a 3-state model might be able to provide the additional flexibility needed here.

\begin{knitrout}
\definecolor{shadecolor}{rgb}{0.969, 0.969, 0.969}\color{fgcolor}\begin{kframe}
\begin{alltt}
\hlcom{# Weights for state-specific density functions}
\hldef{w} \hlkwb{<-} \hlkwd{table}\hldef{(hmm}\hlopt{$}\hlkwd{viterbi}\hldef{())}\hlopt{/}\hlkwd{nrow}\hldef{(data)}

\hlcom{# Select 100 random posterior samples}
\hldef{ind_post} \hlkwb{<-} \hlkwd{sort}\hldef{(}\hlkwd{sample}\hldef{(}\hlnum{1}\hlopt{:}\hldef{(n_iter}\hlopt{/}\hlnum{2}\hldef{),} \hlkwc{size} \hldef{=} \hlnum{100}\hldef{))}

\hlcom{# For each posterior sample, compute gamma distribution}
\hldef{dens_df} \hlkwb{<-} \hlkwd{data.frame}\hldef{()}
\hldef{activity} \hlkwb{<-} \hlkwd{seq}\hldef{(}\hlnum{0.4}\hldef{,} \hlnum{200}\hldef{,} \hlkwc{length} \hldef{=} \hlnum{100}\hldef{)}
\hlkwa{for}\hldef{(i} \hlkwa{in} \hldef{ind_post) \{}
  \hldef{par} \hlkwb{<-} \hldef{hmm}\hlopt{$}\hlkwd{iters}\hldef{()[i,}\hlnum{1}\hlopt{:}\hlnum{6}\hldef{]}
  \hldef{dens1} \hlkwb{<-} \hldef{obs}\hlopt{$}\hlkwd{dists}\hldef{()}\hlopt{$}\hldef{activity}\hlopt{$}\hlkwd{pdf}\hldef{()(}\hlkwc{x} \hldef{= activity,} \hlkwc{mean} \hldef{= par[}\hlnum{1}\hldef{],}
                                      \hlkwc{sd} \hldef{= par[}\hlnum{3}\hldef{],} \hlkwc{z} \hldef{= par[}\hlnum{5}\hldef{])}
  \hldef{dens2} \hlkwb{<-} \hldef{obs}\hlopt{$}\hlkwd{dists}\hldef{()}\hlopt{$}\hldef{activity}\hlopt{$}\hlkwd{pdf}\hldef{()(}\hlkwc{x} \hldef{= activity,} \hlkwc{mean} \hldef{= par[}\hlnum{2}\hldef{],}
                                      \hlkwc{sd} \hldef{= par[}\hlnum{4}\hldef{],} \hlkwc{z} \hldef{= par[}\hlnum{6}\hldef{])}
  \hldef{dens} \hlkwb{<-} \hlkwd{data.frame}\hldef{(}\hlkwc{state} \hldef{=} \hlkwd{paste0}\hldef{(}\hlsng{"state "}\hldef{,} \hlkwd{rep}\hldef{(}\hlnum{1}\hlopt{:}\hlnum{2}\hldef{,} \hlkwc{each} \hldef{=} \hlnum{100}\hldef{)),}
                     \hlkwc{dens} \hldef{=} \hlkwd{c}\hldef{(w[}\hlnum{1}\hldef{]} \hlopt{*} \hldef{dens1, w[}\hlnum{2}\hldef{]} \hlopt{*} \hldef{dens2))}
  \hldef{dens}\hlopt{$}\hldef{group} \hlkwb{<-} \hlkwd{paste0}\hldef{(}\hlsng{"iter "}\hldef{, i,} \hlsng{" - "}\hldef{, dens}\hlopt{$}\hldef{state)}
  \hldef{dens_df} \hlkwb{<-} \hlkwd{rbind}\hldef{(dens_df, dens)}
\hldef{\}}
\hldef{dens_df}\hlopt{$}\hldef{activity} \hlkwb{<-} \hldef{activity}

\hlcom{# Plot histogram of activity and density lines}
\hlkwd{ggplot}\hldef{(dens_df,} \hlkwd{aes}\hldef{(activity, dens))} \hlopt{+}
    \hlkwd{geom_histogram}\hldef{(}\hlkwd{aes}\hldef{(}\hlkwc{y} \hldef{= ..density..),} \hlkwc{data} \hldef{= data,} \hlkwc{fill} \hldef{=} \hlsng{"lightgrey"}\hldef{,}
                   \hlkwc{col} \hldef{=} \hlsng{"grey"}\hldef{,} \hlkwc{breaks} \hldef{=} \hlkwd{seq}\hldef{(}\hlnum{0}\hldef{,} \hlnum{200}\hldef{,} \hlkwc{by} \hldef{=} \hlnum{10}\hldef{))} \hlopt{+}
    \hlkwd{geom_line}\hldef{(}\hlkwd{aes}\hldef{(}\hlkwc{col} \hldef{= state,} \hlkwc{group} \hldef{= group),} \hlkwc{size} \hldef{=} \hlnum{0.1}\hldef{,} \hlkwc{alpha} \hldef{=} \hlnum{0.5}\hldef{)} \hlopt{+}
    \hlkwd{scale_color_manual}\hldef{(}\hlkwc{values} \hldef{= hmmTMB}\hlopt{:::}\hldef{hmmTMB_cols,} \hlkwc{name} \hldef{=} \hlkwa{NULL}\hldef{)} \hlopt{+}
    \hlkwd{guides}\hldef{(}\hlkwc{color} \hldef{=} \hlkwd{guide_legend}\hldef{(}\hlkwc{override.aes} \hldef{=} \hlkwd{list}\hldef{(}\hlkwc{size} \hldef{=} \hlnum{0.5}\hldef{,} \hlkwc{alpha} \hldef{=} \hlnum{1}\hldef{)))}
\end{alltt}
\end{kframe}

{\centering \includegraphics[width=0.8\linewidth]{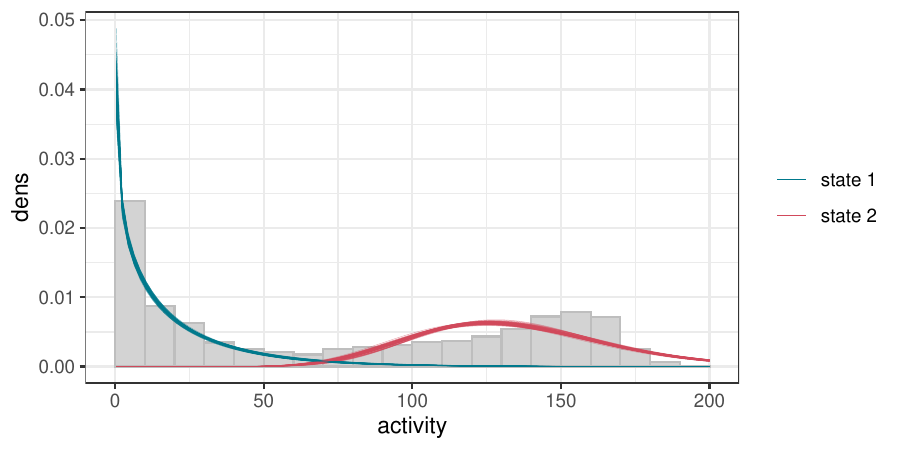} 

}

\end{knitrout}

\subsection{Covariate effects}

The aim of the study was to estimate daily variations in activity. We can use the function \code{HMM\$predict()} to compute the stationary state probabilities over a grid of values of time of day. The stationary state probabilities are the stationary distribution of the corresponding transition probabilities, and they are meant to roughly measure the proportion of time spent in each state, for a given covariate value. Like above, we generate curves from 100 randomly selected posterior samples, so that the plot captures the uncertainty in the estimated relationship. To do this, we use \code{hmm\$update\_par(iter = i)} to update the parameter values stored in the model (and used in \code{HMM\$predict()}) to the $i$-th posterior sample, and we loop over $i$. 

Although there is high variation between the posterior samples, they almost all indicate that it was much more likely to be in the high activity state during the day, and in the low activity state at night.

\begin{knitrout}
\definecolor{shadecolor}{rgb}{0.969, 0.969, 0.969}\color{fgcolor}\begin{kframe}
\begin{alltt}
\hlcom{# Select 100 random posterior samples}
\hldef{ind_post} \hlkwb{<-} \hlkwd{sort}\hldef{(}\hlkwd{sample}\hldef{(}\hlnum{1}\hlopt{:}\hldef{(n_iter}\hlopt{/}\hlnum{2}\hldef{),} \hlkwc{size} \hldef{=} \hlnum{100}\hldef{))}

\hlcom{# For each posterior sample, compute stationary state probabilities over}
\hlcom{# grid of time of day}
\hldef{probs_df} \hlkwb{<-} \hlkwd{data.frame}\hldef{()}
\hldef{newdata} \hlkwb{<-} \hlkwd{data.frame}\hldef{(}\hlkwc{tod} \hldef{=} \hlkwd{seq}\hldef{(}\hlnum{0}\hldef{,} \hlnum{24}\hldef{,} \hlkwc{length} \hldef{=} \hlnum{100}\hldef{))}
\hlkwa{for}\hldef{(i} \hlkwa{in} \hldef{ind_post) \{}
  \hldef{hmm}\hlopt{$}\hlkwd{update_par}\hldef{(}\hlkwc{iter} \hldef{= i)}
  \hldef{probs} \hlkwb{<-} \hlkwd{data.frame}\hldef{(}\hlkwc{state} \hldef{=} \hlkwd{rep}\hldef{(}\hlkwd{paste0}\hldef{(}\hlsng{"state "}\hldef{,} \hlnum{1}\hlopt{:}\hlnum{2}\hldef{),} \hlkwc{each} \hldef{=} \hlnum{100}\hldef{),}
                      \hlkwc{prob} \hldef{=} \hlkwd{as.vector}\hldef{(hmm}\hlopt{$}\hlkwd{predict}\hldef{(}\hlkwc{what} \hldef{=} \hlsng{"delta"}\hldef{,}
                                                   \hlkwc{newdata} \hldef{= newdata)))}
  \hldef{probs}\hlopt{$}\hldef{group} \hlkwb{<-} \hlkwd{paste0}\hldef{(}\hlsng{"iter "}\hldef{, i,} \hlsng{" - "}\hldef{, probs}\hlopt{$}\hldef{state)}
  \hldef{probs_df} \hlkwb{<-} \hlkwd{rbind}\hldef{(probs_df, probs)}
\hldef{\}}
\hldef{probs_df}\hlopt{$}\hldef{tod} \hlkwb{<-} \hldef{newdata}\hlopt{$}\hldef{tod}

\hlcom{# Plot stationary state probs against time of day}
\hlkwd{ggplot}\hldef{(probs_df,} \hlkwd{aes}\hldef{(tod, prob,} \hlkwc{group} \hldef{= group,} \hlkwc{col} \hldef{= state))} \hlopt{+}
  \hlkwd{geom_line}\hldef{(}\hlkwc{size} \hldef{=} \hlnum{0.1}\hldef{,} \hlkwc{alpha} \hldef{=} \hlnum{0.5}\hldef{)} \hlopt{+}
  \hlkwd{scale_x_continuous}\hldef{(}\hlkwc{breaks} \hldef{=} \hlkwd{seq}\hldef{(}\hlnum{0}\hldef{,} \hlnum{24}\hldef{,} \hlkwc{by} \hldef{=} \hlnum{4}\hldef{))} \hlopt{+}
  \hlkwd{labs}\hldef{(}\hlkwc{x} \hldef{=} \hlsng{"time of day"}\hldef{,} \hlkwc{y} \hldef{=} \hlsng{"stationary state probabilities"}\hldef{,} \hlkwc{col} \hldef{=} \hlkwa{NULL}\hldef{)} \hlopt{+}
  \hlkwd{scale_color_manual}\hldef{(}\hlkwc{values} \hldef{= hmmTMB}\hlopt{:::}\hldef{hmmTMB_cols)} \hlopt{+}
  \hlkwd{guides}\hldef{(}\hlkwc{color} \hldef{=} \hlkwd{guide_legend}\hldef{(}\hlkwc{override.aes} \hldef{=} \hlkwd{list}\hldef{(}\hlkwc{size} \hldef{=} \hlnum{0.5}\hldef{,} \hlkwc{alpha} \hldef{=} \hlnum{1}\hldef{)))}
\end{alltt}
\end{kframe}

{\centering \includegraphics[width=0.8\linewidth]{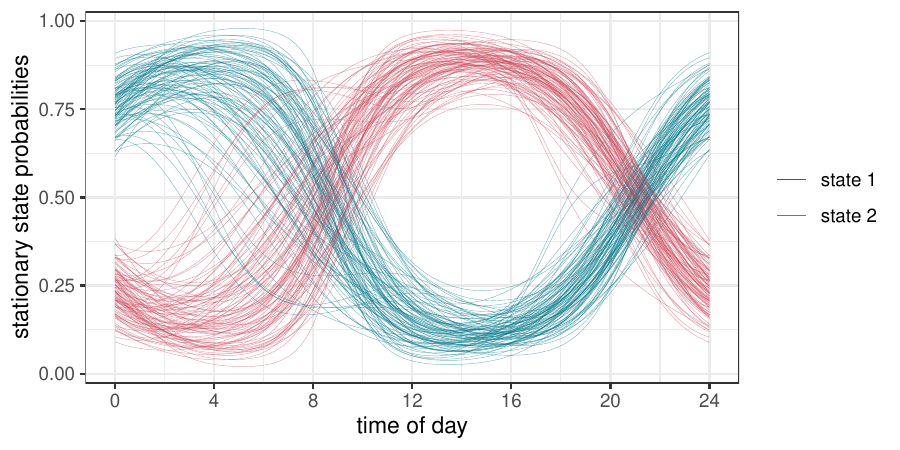} 

}

\end{knitrout}

\section{Simulation study}
\label{app:sim}

In this appendix, we investigate the ability of \pkg{hmmTMB} to recover relationships between parameters and covariates in three scenarios: (1) non-homogeneous hidden Markov model with non-linear covariate effects on transition probabilities, (2) Markov-switching generalised additive model with non-linear covariate effects on observation parameters, and (3) mixed hidden Markov model with random effect on transition probabilities.

\subsection{Non-homogeneous hidden Markov model}

We first considered a 2-state HMM where the transition probabilities are functions of a time-varying continuous covariate $x_{1t} \in [-1, 1]$. The true relationship used to simulate data was
\begin{align*}
  \text{logit}(\gamma_{12}^{(t)}) & = -3 + 3 x_{1t}^2 \\
  \text{logit}(\gamma_{21}^{(t)}) & = -2 + \sin(\pi x_{1t}).
\end{align*}

The observations were simulated from state-dependent normal distributions,
\begin{equation*}
  Z_t \mid \{ S_t = j \} \sim N(\mu_j, \sigma_j^2)
\end{equation*}
where $\mu_1 = -5$, $\mu = 5$, and $\sigma_1 = \sigma_2 = 1$.

We fitted a 2-state HMM where the transition probabilities were modelled with non-parametric splines. We used the \code{"cs"} basis (cubic spline) for $\gamma_{12}^{(t)}$ and the \code{"cc"} basis (cyclical cubic spline) for $\gamma_{21}^{(t)}$. We assumed normal observation distributions, for which the parameters were estimated during model fitting.

We repeated the procedure 200 times; at each iteration,
\begin{enumerate}
\item Simulate $n = 5000$ values of the covariate using a reflected Gaussian random walk on $(-1, 1)$.
\item Derive one transition probability matrix for each time step $t = 1, \dots, n$, based on the assumed relationships.
\item Simulate state sequence based on transition matrices, and then simulate observations $z_1, \dots, z_n$ from state-dependent distributions.
\item Fit HMM to the observations $z_1, \dots, z_n$.
\item Predict the transition probabilities over a grid of values of $x_1$.
\end{enumerate}

\begin{figure}[htbp]
  \centering
  \includegraphics[width=0.49\textwidth]{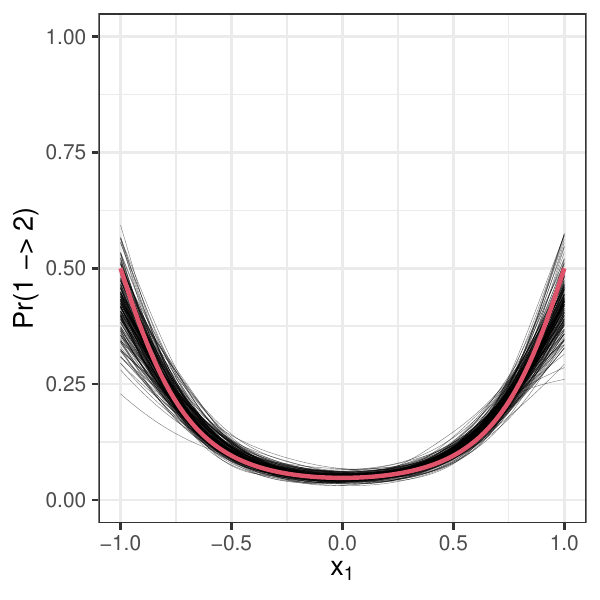}
  \includegraphics[width=0.49\textwidth]{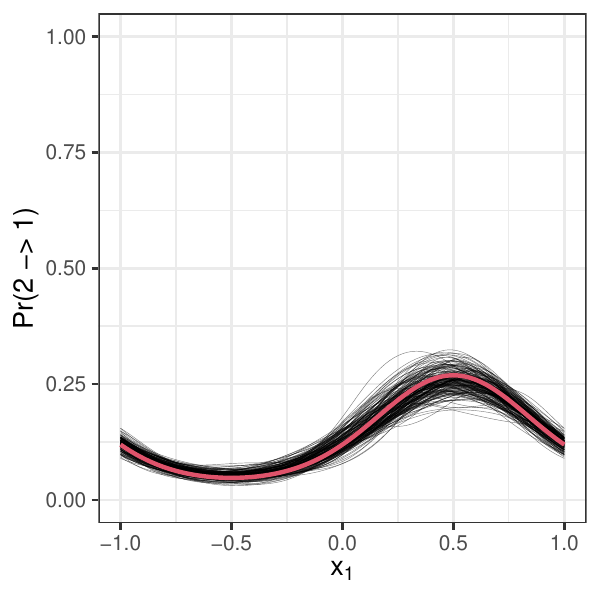}
  \caption{Results of non-homogeneous HMM simulation. The red lines are the true relationships between transition probabilities and the covariate $x_1$ used for simulating, and the black lines are estimates.}
  \label{fig:sim1}
\end{figure}

The estimated relationships between the transition probabilities and the covariate are shown in Figure \ref{fig:sim1}, together with the true relationships. For each transition probability, all estimates captured the shape of the true function well.

\subsection{Markov-switching regression}

Next, we considered a 2-state HMM with covariate effects on the observation parameters. Specifically, we modelled observations with state-dependent Poisson distributions,
\begin{equation*}
  Z_t \mid \{ S_t = j \} \sim \text{Poisson}(\lambda_j^{(t)}).
\end{equation*}

The rate parameter was chosen to be constant in state 2 (i.e., $\lambda_2^{(t)} = \lambda_2$), and to be a function of a continuous time-varying covariate $x_{1t} \in [-1, 1]$ in state 1:
\begin{equation*}
  \lambda_1^{(t)} =
  \begin{cases}
    \exp(1 + \sin(2 \pi x_{1t})) & \text{if } -0.5 \leq x_{1t} \leq 0.5,\\
    \exp(1) & \text{otherwise.}
  \end{cases}
\end{equation*}

We tried to recover this relationship using a Markov-switching generalised additive model, i.e., by modelling the (log) rate with a cubic spline (\code{"cs"} in \pkg{mgcv} syntax). 

We repeated the procedure 200 times; at each iteration,
\begin{enumerate}
\item Simulate $n = 2000$ values of the covariate using a reflected Gaussian random walk on $(-1, 1)$.
\item Derive the time-varying Poisson rates based on the function described above.
\item Simulate state sequence based on the transition probabilities $\gamma_{12} = \gamma_{21} = 0.1$, and then simulate observations $z_1, \dots, z_n$ from state-dependent distributions with time-varying parameters.
\item Fit HMM to the observations $z_1, \dots, z_n$.
\item Predict the Poisson rate in state 1 over a grid of values of $x_1$.
\end{enumerate}

\begin{figure}[htbp]
  \centering
  \includegraphics[width=0.7\textwidth]{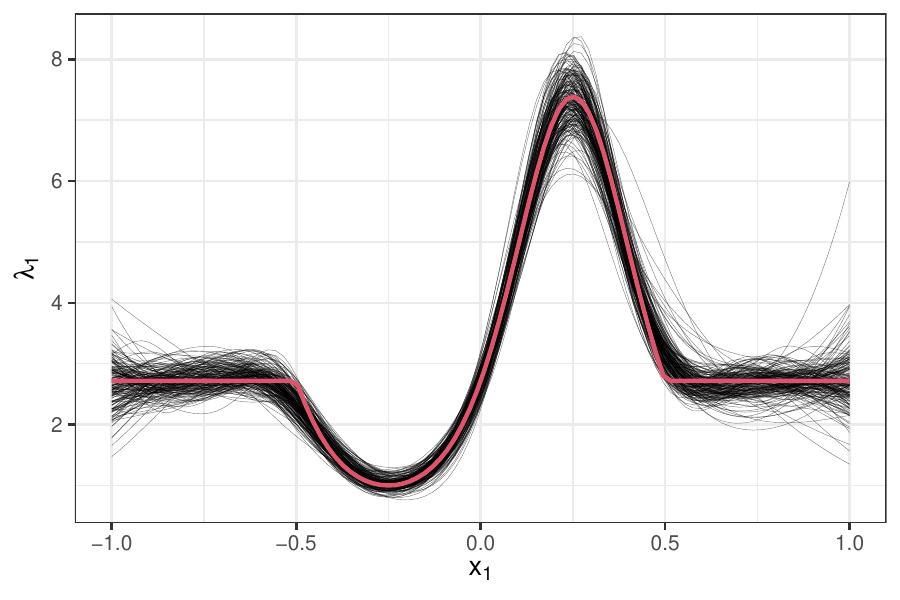}
  \caption{Results of Markov-switching regression simulation. The red line is the true relationship between the rate parameter $\lambda_1$ and the covariate $x_1$ used in the simulations, and the black lines are estimates.}
  \label{fig:sim2}
\end{figure}

The estimated parameters $\lambda_1$ are shown as functions of $x_1$ in Figure \ref{fig:sim2}. The true non-linear function was captured well by the estimated splines.

\subsection{Mixed hidden Markov model}

We simulated data from a 2-state HMM where the transition probabilities were themselves random, to check whether their distribution could be estimated using \pkg{hmmTMB}. We consider individual-specific transition probabilities (where the individual could be a subject in the study, or an animal, or any relevant grouping) where, for individual $m \in \{ 1, \dots, M \}$,
\begin{align*}
  \text{logit}(\gamma_{12}^{(m)}) & = -2.5 + \beta_{12}^{(m)},\quad \beta_{12}^{(m)} \sim N(0, 1^2) \\
  \text{logit}(\gamma_{21}^{(m)}) & = -2.5 + \beta_{21}^{(m)},\quad \beta_{21}^{(m)} \sim N(0, 0.5^2).
\end{align*}

In this scenario, the aim was to estimate the variance of the distributions of the random effects $\beta_{12}^{(m)}$ and $\beta_{21}^{(m)}$.

Observations were simulated from state-dependent gamma distributions,
\begin{equation*}
  Z_t \mid \{ S_t = j \} \sim \text{gamma}(\mu_j, \sigma_j^2), 
\end{equation*}
with means $\mu_1 = 3$ and $\mu_2 = 15$, and standard deviations $\sigma_1 = 2$ and $\sigma_2 = 5$.

We fitted an HMM where the transition probabilities included an i.i.d.\ normal random effect for the individual, using \code{s(ID, bs = "re")}.

We repeated the procedure 200 times; at each iteration,
\begin{enumerate}
\item Randomly generate 20 transition probability matrices (one for each of 20 individuals), based on the true distribution described above. 
\item For each individual, simulate a state sequence of length 500 based on the individual-specific transition probability matrix, and then simulate observations from the state-dependent gamma distributions.
\item Fit HMM to the observations from all individuals.
\item Save the estimated standard deviation of the random effects.
\end{enumerate}

\begin{figure}[htbp]
  \centering
  \includegraphics[width=0.7\textwidth]{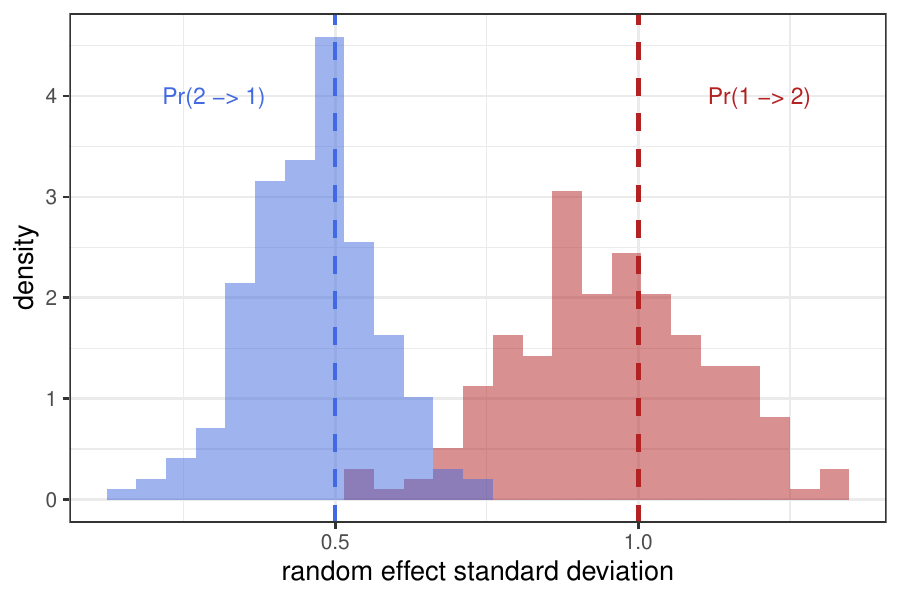}
  \caption{Results of mixed HMM simulation. The vertical dotted lines show the true values of the standard deviation of random effects used in the simulations, and the histograms show the distribution of 200 estimates.}
  \label{fig:sim3}
\end{figure}

The distributions of the estimated random effect standard deviations are shown in Figure \ref{fig:sim3}. The true standard deviations were recovered successfully in most analyses.

\section{Petrel data preparation code}
\label{app:petrel}

The code below loads the data from \cite{descamps2016data} from the Movebank data repository, and prepares it for the HMM analysis. In particular, it:
\begin{itemize}
\item removes unneeded columns from the original data;
\item keeps only 10 tracks;
\item computes distance to centre for each track;
\item computes step lengths and turning angles.
\end{itemize}

\begin{knitrout}
\definecolor{shadecolor}{rgb}{0.969, 0.969, 0.969}\color{fgcolor}\begin{kframe}
\begin{alltt}
\hlkwd{library}\hldef{(moveHMM)}
\hlkwd{library}\hldef{(sp)}

\hlcom{# Load data from Movebank}
\hldef{URL} \hlkwb{<-} \hlkwd{paste0}\hldef{(}\hlsng{"https://www.datarepository.movebank.org/bitstream/handle/10255/"}\hldef{,}
              \hlsng{"move.568/At-sea%20distribution%20Antarctic%20Petrel%2c%"}\hldef{,}
              \hlsng{"20Antarctica%202012%20%28data%20from%20Descamps%20et%20al.%"}\hldef{,}
              \hlsng{"202016%29-gps.csv"}\hldef{)}
\hldef{raw} \hlkwb{<-} \hlkwd{read.csv}\hldef{(}\hlkwd{url}\hldef{(URL))}

\hlcom{# Keep relevant columns: ID, time, lon, lat}
\hldef{data_all} \hlkwb{<-} \hldef{raw[,} \hlkwd{c}\hldef{(}\hlnum{13}\hldef{,} \hlnum{3}\hldef{,} \hlnum{4}\hldef{,} \hlnum{5}\hldef{)]}
\hlkwd{colnames}\hldef{(data_all)} \hlkwb{<-} \hlkwd{c}\hldef{(}\hlsng{"ID"}\hldef{,} \hlsng{"time"}\hldef{,} \hlsng{"lon"}\hldef{,} \hlsng{"lat"}\hldef{)}
\hldef{data_all}\hlopt{$}\hldef{time} \hlkwb{<-} \hlkwd{as.POSIXct}\hldef{(data_all}\hlopt{$}\hldef{time,} \hlkwc{tz} \hldef{=} \hlsng{"MST"}\hldef{)}

\hlcom{# Function to compute first-order differences for grouped data}
\hldef{diff_by_ID} \hlkwb{<-} \hlkwa{function}\hldef{(}\hlkwc{x}\hldef{,} \hlkwc{ID}\hldef{,} \hlkwc{...}\hldef{) \{}
    \hlcom{# Indices of first and last value of each group}
    \hldef{n} \hlkwb{<-} \hlkwd{length}\hldef{(ID)}
    \hldef{i0} \hlkwb{<-} \hlkwd{which}\hldef{(ID[}\hlopt{-}\hlnum{1}\hldef{]} \hlopt{!=} \hldef{ID[}\hlopt{-}\hldef{n])}
    \hldef{i_first} \hlkwb{<-} \hlkwd{c}\hldef{(}\hlnum{1}\hldef{, i0} \hlopt{+} \hlnum{1}\hldef{)}
    \hldef{i_last} \hlkwb{<-} \hlkwd{c}\hldef{(i0, n)}

    \hlcom{# First-order differences}
    \hldef{dx} \hlkwb{<-} \hlkwd{rep}\hldef{(}\hlnum{NA}\hldef{, n)}
    \hldef{dx[}\hlopt{-}\hldef{i_last]} \hlkwb{<-} \hlkwd{difftime}\hldef{(}\hlkwc{time1} \hldef{= x[}\hlopt{-}\hldef{i_first],} \hlkwc{time2} \hldef{= x[}\hlopt{-}\hldef{i_last], ...)}
    \hlkwd{return}\hldef{(dx)}
\hldef{\}}

\hlcom{# Keep only a few tracks for this example (excluding tracks that}
\hlcom{# have unusually long intervals)}
\hldef{dtimes} \hlkwb{<-} \hlkwd{diff_by_ID}\hldef{(data_all}\hlopt{$}\hldef{time, data_all}\hlopt{$}\hldef{ID)}
\hldef{keep_ids} \hlkwb{<-} \hlkwd{setdiff}\hldef{(}\hlkwd{unique}\hldef{(data_all}\hlopt{$}\hldef{ID),}
                    \hlkwd{unique}\hldef{(data_all}\hlopt{$}\hldef{ID[}\hlkwd{which}\hldef{(dtimes} \hlopt{>} \hlnum{30}\hldef{)]))[}\hlnum{1}\hlopt{:}\hlnum{10}\hldef{]}
\hldef{data} \hlkwb{<-} \hlkwd{subset}\hldef{(data_all, ID} \hlopt{%in%} \hldef{keep_ids)}
\hldef{data} \hlkwb{<-} \hldef{data[}\hlkwd{with}\hldef{(data,} \hlkwd{order}\hldef{(ID, time)),]}

\hlcom{# Define centre for each track as first observation}
\hldef{i0} \hlkwb{<-} \hlkwd{c}\hldef{(}\hlnum{1}\hldef{,} \hlkwd{which}\hldef{(data}\hlopt{$}\hldef{ID[}\hlopt{-}\hlnum{1}\hldef{]} \hlopt{!=} \hldef{data}\hlopt{$}\hldef{ID[}\hlopt{-}\hlkwd{nrow}\hldef{(data)])} \hlopt{+} \hlnum{1}\hldef{)}
\hldef{centres} \hlkwb{<-} \hldef{data[i0,} \hlkwd{c}\hldef{(}\hlsng{"ID"}\hldef{,} \hlsng{"lon"}\hldef{,} \hlsng{"lat"}\hldef{)]}
\hldef{data}\hlopt{$}\hldef{centre_lon} \hlkwb{<-} \hlkwd{rep}\hldef{(centres}\hlopt{$}\hldef{lon,} \hlkwd{rle}\hldef{(data}\hlopt{$}\hldef{ID)}\hlopt{$}\hldef{lengths)}
\hldef{data}\hlopt{$}\hldef{centre_lat} \hlkwb{<-} \hlkwd{rep}\hldef{(centres}\hlopt{$}\hldef{lat,} \hlkwd{rle}\hldef{(data}\hlopt{$}\hldef{ID)}\hlopt{$}\hldef{lengths)}

\hlcom{# Add distance to centre as covariate (based on sp for great circle distance)}
\hldef{data}\hlopt{$}\hldef{d2c} \hlkwb{<-} \hlkwd{sapply}\hldef{(}\hlnum{1}\hlopt{:}\hlkwd{nrow}\hldef{(data),} \hlkwa{function}\hldef{(}\hlkwc{i}\hldef{) \{}
    \hlkwd{spDistsN1}\hldef{(}\hlkwc{pts} \hldef{=} \hlkwd{matrix}\hldef{(}\hlkwd{as.numeric}\hldef{(data[i,} \hlkwd{c}\hldef{(}\hlsng{"lon"}\hldef{,} \hlsng{"lat"}\hldef{)]),} \hlkwc{ncol} \hldef{=} \hlnum{2}\hldef{),}
              \hlkwc{pt} \hldef{=} \hlkwd{c}\hldef{(data}\hlopt{$}\hldef{centre_lon[i], data}\hlopt{$}\hldef{centre_lat[i]),}
              \hlkwc{longlat} \hldef{=} \hlnum{TRUE}\hldef{)}
\hldef{\})}
\hlcom{# Remove unnecessary columns}
\hldef{data}\hlopt{$}\hldef{centre_lon} \hlkwb{<-} \hlkwa{NULL}
\hldef{data}\hlopt{$}\hldef{centre_lat} \hlkwb{<-} \hlkwa{NULL}

\hlcom{# Derive step lengths and turning angles using moveHMM}
\hldef{movehmm_data} \hlkwb{<-} \hlkwd{prepData}\hldef{(}\hlkwc{trackData} \hldef{= data,}
                         \hlkwc{coordNames} \hldef{=} \hlkwd{c}\hldef{(}\hlsng{"lon"}\hldef{,} \hlsng{"lat"}\hldef{),}
                         \hlkwc{type} \hldef{=} \hlsng{"LL"}\hldef{)}
\hldef{data}\hlopt{$}\hldef{step} \hlkwb{<-} \hldef{movehmm_data}\hlopt{$}\hldef{step}
\hldef{data}\hlopt{$}\hldef{angle} \hlkwb{<-} \hldef{movehmm_data}\hlopt{$}\hldef{angle}

\hlcom{# Replace zero step length to very small number because it's overkill}
\hlcom{# to use a zero-inflated distribution just for one zero observation}
\hldef{wh_zero} \hlkwb{<-} \hlkwd{which}\hldef{(data}\hlopt{$}\hldef{step} \hlopt{==} \hlnum{0}\hldef{)}
\hldef{data}\hlopt{$}\hldef{step[wh_zero]} \hlkwb{<-} \hlkwd{runif}\hldef{(}\hlkwd{length}\hldef{(wh_zero),}
                            \hlkwc{min} \hldef{=} \hlnum{0}\hldef{,}
                            \hlkwc{max} \hldef{=} \hlkwd{min}\hldef{(data}\hlopt{$}\hldef{step[}\hlopt{-}\hldef{wh_zero],} \hlkwc{na.rm} \hldef{=} \hlnum{TRUE}\hldef{))}

\hlcom{# Shorten track names}
\hldef{data}\hlopt{$}\hldef{ID} \hlkwb{<-} \hlkwd{factor}\hldef{(data}\hlopt{$}\hldef{ID)}
\hlkwd{levels}\hldef{(data}\hlopt{$}\hldef{ID)} \hlkwb{<-} \hlkwd{paste0}\hldef{(}\hlsng{"PET-"}\hldef{, LETTERS[}\hlnum{1}\hlopt{:}\hlkwd{length}\hldef{(}\hlkwd{unique}\hldef{(data}\hlopt{$}\hldef{ID))])}

\hlkwd{write.csv}\hldef{(data,} \hlkwc{file} \hldef{=} \hlsng{"petrels.csv"}\hldef{,} \hlkwc{row.names} \hldef{=} \hlnum{FALSE}\hldef{)}
\end{alltt}
\end{kframe}
\end{knitrout}

\end{appendix}

\end{document}